\documentclass[modern]{aastex63}

\newcommand\luh{\object{Luhman\,16}}

\received{April 21, 2020}
\revised{??}
\accepted{\today}

\submitjournal{AJ}

\shorttitle{TESS Lightcurve of Luhman\,16}
\shortauthors{Apai et al.}
\graphicspath{{./}{figures/}}

\begin{document}


\title{{\it TESS} Observations of the Luhman\,16\,AB Brown Dwarf System: Rotational Periods, Lightcurve Evolution, and Zonal Circulation \footnote{Based on observations collected with the NASA {\it TESS}.}}

\correspondingauthor{D\'aniel Apai}
\email{apai@arizona.edu}

\author[0000-0003-3714-5855]{D\'aniel Apai}
\affiliation{Steward Observatory, The University of Arizona\\
933 N. Cherry Avenue, Tucson, AZ 85721, USA}
\affiliation{Lunar and Planetary Laboratory, The University of Arizona,\\ 1629 E. University. Blvd., Tucson, AZ 85721, USA}

\author[0000-0003-1149-3659]{Domenico Nardiello}
\affiliation{Aix Marseille Univ, CNRS, CNES, LAM, Marseille, France}
\affiliation{INAF - Osservatorio Astronomico di Padova, Vicolo dell'Osservatorio 5, IT-35122, Padova, Italy}

\author[0000-0003-4080-6466]{Luigi R. Bedin}
\affiliation{INAF - Osservatorio Astronomico di Padova, Vicolo dell'Osservatorio 5, IT-35122, Padova, Italy}




\begin{abstract}
Brown dwarfs were recently found to display rotational modulations, commonly attributed to cloud cover of varying thickness, possibly modulated by planetary-scale waves. However, the long-term, continuous, high-precision monitoring data to test this hypothesis for more objects is lacking. By applying our novel photometric approach to TESS data, we extract a high-precision lightcurve of the closest brown dwarfs, which form the binary system Luhman 16AB. Our observations, that cover about 100 rotations of Luhman 16B, display continuous lightcurve evolution. The periodogram analysis shows that the rotational period of the component that dominates the lightcurve is 5.28 h. We also find evidence for periods of 2.5 h, 6.94 h, and 90.8 h. We show that the 2.5 h and 5.28 h periods emerge from Luhman 16B and that they consist of multiple, slightly shifted peaks, revealing the presence of high-speed jets and zonal circulation in this object. We find that the lightcurve evolution is well fit by the planetary-scale waves model, further supporting this interpretation. We argue that the 6.94 h peak is likely the rotation period of Luhman 16A. By comparing the rotational periods to observed v sin(i) measurements, we show that the two brown dwarfs are viewed at angles close to their equatorial planes. We also describe a long-period (P$\sim$91 h) evolution in the lightcurve, which we propose emerges from the vortex-dominated polar regions. Our study paves the way toward direct comparisons of the predictions of global circulation models to observations via periodogram analysis.
\end{abstract}

\keywords{stars: atmospheres --- brown dwarfs --- Jupiter --- planets and satellites: atmospheres --- planets and satellites: gaseous planets}


\section{Introduction} 
\label{introduction}
%

Planetary and brown dwarf atmospheres are fundamentally impacted by the presence of condensate clouds \citep[e.g.,][]{Burrows2001,RobinsonMarley2014,Zhang2020}, and several studies showed that the properties of clouds themselves are impacted by atmosphere dynamics \citep[e.g.,][]{ShowmanKaspi2012,ShowmanKaspi2013,Showman2020}. In weakly irradiated  atmospheres (such as directly imaged exoplanets and brown dwarfs) the interplay of rotation and heat transfer will set the nature of large-scale atmospheric circulation \citep[e.g.,][]{ShowmanKaspi2012,ZhangShowman2014}, The presence or absence of clouds will impact the local pressure-temperature profiles which, in turn, can lead to the formation/evaporative dissipation of clouds \citep[][]{Tan2017}, triggering self-sustaining cycles of cloud formation and, possibly, driving planetary-scale waves. Atmospheric layers can also be heated by breaking of atmospheric waves emerging from high-pressure turbulent layers, generating planetary-scale waves \citep[][]{Tan2019}.

The presence of planetary-scale waves has recently been invoked to explain data from long-term monitoring of L/T spectral type transition brown dwarfs \citep[][]{Apai2017}. In three brown dwarfs (two of which are planetary-mass), \citet[][]{Apai2017} observed continuous lightcurve evolution over timescales of 1--1,000 rotational periods. They showed that bright or dark features alone (i.e., one to few Great Red Spot-like features) cannot explain the symmetric, wave-like lightcurves (as spots tend to introduce asymmetric variations). Instead, \citet{Apai2017} demonstrated that planetary-scale waves can provide a simple explanation for the observed lightcurve evolution -- a simple model of a sum of two- to three sine waves (with unconstrained amplitude, phase, and period) successfully fitted all lightcurve segments. This study raised two questions: A) Are brown dwarfs other than the three studied exhibit lightcurve evolution that is consistent with the planetary-scale model? and B) What processes drive the planetary-scale waves? Answering these two questions requires long-term, high-precision, photometric monitoring of a larger sample of brown dwarfs, but the end of the Spitzer Space Telescope mission led to the loss of continuous infrared monitoring capabilities. The study presented here demonstrates a novel approach to obtain the required data.

Although the studies described focus on brown dwarfs, it is paramount for the reader to understand that brown dwarfs -- including planetary-mass objects -- also provide a gateway to understanding  directly imaged exoplanets. In fact, L/T spectral type transition brown dwarfs are excellent analogs to the directly imaged exoplanets in the HR~8799 system \citep[][]{Marois2010}.

This study focuses on the L/T spectral transition brown dwarf binary \object{WISEJ104915.57-531906.1} or, in the following, \luh{}. 
Even though the closest known brown dwarf system to the Sun (d$=1.9955\pm0.0004$\,pc, \citealt{Bedin2017}), \luh{} was discovered only in 2013 \citep{Luhman2013}. The binary consists of a  34.2$^{+1.3}_{-1.1}$~M$_{Jup}$ L-type primary and a 27.9$^{+1.0}_{-1.1}$~M$_{Jup}$ T-type secondary brown dwarf component \citep[][]{Ammons2019}. The primary's spectral type is L7.5 and the secondary's spectral type is T0.5 \citep[][]{Kniazev2013,Burgasser2013}.
Right after its discovery \luh{} was identified as one of the most highly-variable brown dwarfs \citep[][]{Gillon2013,Biller2013} and follow-up spectroscopic observations showed spectrophotometric modulations \citep[][]{Burgasser2014} that resembled those of the few other L/T transition brown dwarfs studied with similar methods \citep[][]{Apai2013}. High-precision spectrophotometric observations in combination with high spatial resolution revealed that \textit{both} A and B components are variable \citep{Biller2013}, although the integrated modulations are dominated by the modulations in the T0.5-type B component of the system and not by the L7.5-type A component \citep[e.g.,][]{Burgasser2014,Crossfield2014,Buenzli2015a}. Detailed modeling of the spectrophotometric modulations showed that spatially correlated cloud thickness--temperature modulations (invoked for other L/T transition brown dwarfs, see \citealt{Radigan2012,Apai2013}) can also explain the behavior of \luh{} \citep{Buenzli2015a,Buenzli2015b}. \luh{} is one of the relatively small number of brown dwarfs bright enough to allow time-resolved medium- and high-resolution spectroscopy that can probe different atomic and molecular tracers to further constrain the properties of the heterogeneous cloud cover. \citet{Buenzli2015b} reported that FeH absorption is very strongly correlated with the continuum brightness, possibly arguing for a lack of deep opacity holes that would allow sightlines into the hotter interior where FeH would be present in gas phase. \citet{Crossfield2014} used Doppler imaging CO inversion technique to present a possible map of the cloud pattern{, but one which is not sensitive to potential longitudinal banded structures}. \citet{Kellogg2017} found that the K~I line is inversely correlated with the continuum brightness variations, suggesting that the cloud thickness variations significantly change the pressure-temperature profile in the upper atmosphere.

{In a new study \citet{Millar-Blanchaer2020} published novel H--band, spatially resolved polarimetric study of the \luh{} system. This study measured a $0.031\%\pm0.004$\% polarization level for Luhman~16\,A and $0.010\%\pm0.004\%$ polarization level for Luhman~16\,B. Detailed modeling showed that while the lower-level polarization of Luhman~16\,B can be explain by either oblateness or by banded cloud structures, Luhman~16\,A's stronger polarization is only consistent with the presence of banded clouds and not with oblateness alone. The \citet{Millar-Blanchaer2020} study assumed a rotation period of 5~h or 8~h for Luhman~16\,A and 5~h for Luhman~16\,B.}

Beyond the spectrophotometric information that can shed light onto the cloud properties, the most perplexing property of \luh{}B is the prominent evolution of its lightcurve. Modeling of multi-epoch lightcurve \textit{snapshots} (covering typically 1 rotational periods) identified  recurring lightcurve feature and were able to explain the lightcurves with few large spots \citep[][]{Karalidi2016}. However, the study of \citet[][]{Apai2017} showed that while single-rotation lightcurves can often be explained with spots or waves, lightcurve evolution observed in brown dwarfs \textit{over multiple rotations} could not be explained by spots and required planetary-scale waves. As a nearby, bright, and highly variable L/T transition brown dwarf, \luh{} provides the ideal target to test the planetary-scale model.

In this paper we present a unique, long-term monitoring observation of \luh{} with NASA's TESS \citep[][]{Ricker2014} satellite, covering about 100 rotational periods. In the following we review the observations and data reduction, we present our lightcurve analysis, and review and discuss the results. Finally, we summarize the conclusions of our study.

%
\section{Observations and Data Reduction}
%

\begin{figure}
\begin{center}
\includegraphics[width=\textwidth]{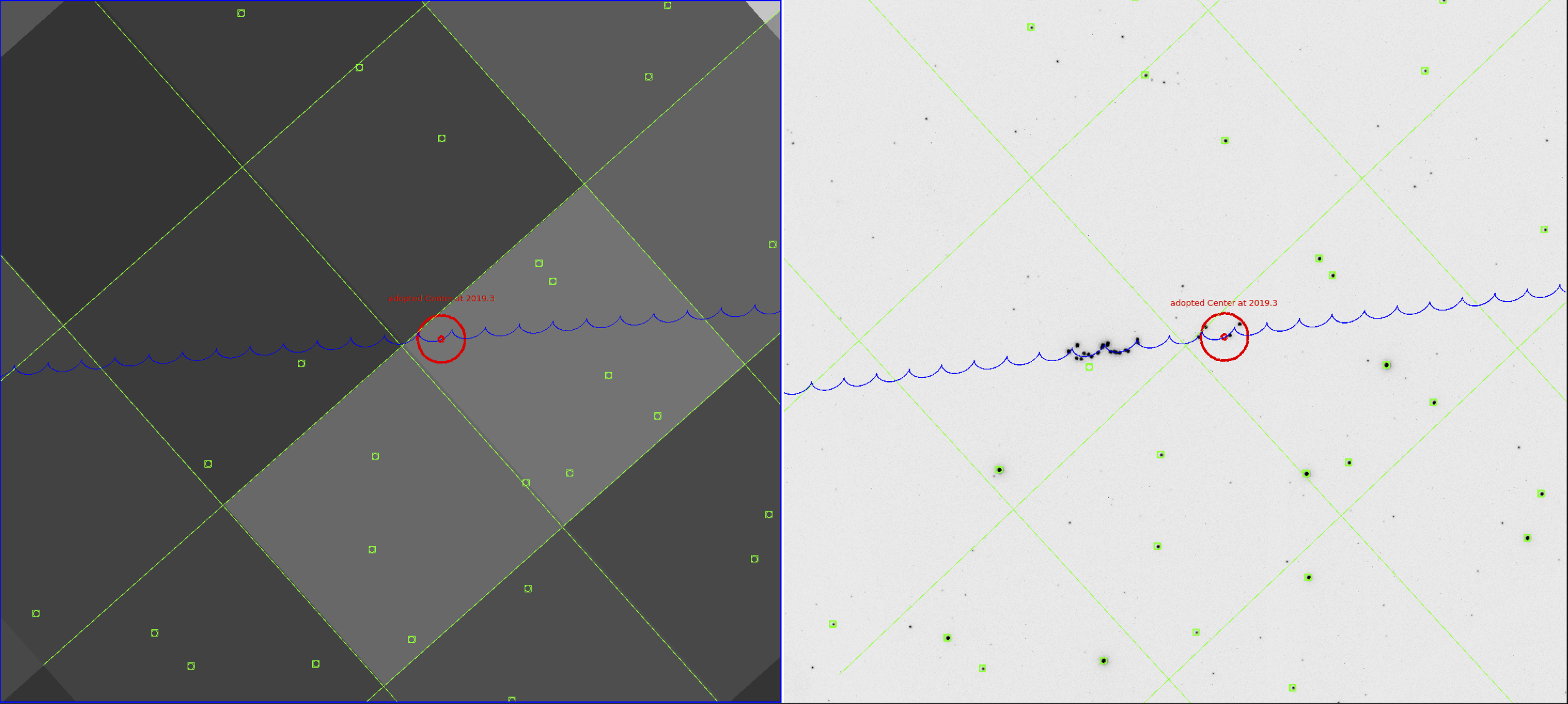}
\caption{\textit{(Left):} Finding chart of Luh\,16\,AB on \textit{TESS} image (\texttt{tess2019107105933}). 
Axes are aligned North up, East to the left, and the FoV is about $66^{\prime\prime}\times 60^{\prime\prime}$.  
\textit{(Right):} Same field of view from \textit{HST} (from \citealt{Bedin2017}). 
Green lines indicate the pixel grid on the \textit{TESS} images. The blue lines is the astrometric solution 
for the baricenter of the Luh\,16\,AB system. The red circles highlight the adopted position for the baricenter at the epoch 
2019.26 during which \textit{TESS} observations were acquired, the smallest of which has a radius of 220\,mas (indicative of the smallest 
separation between the two components).  
The green squares are the sources in the \textit{Gaia\,DR2} catalog. \label{fig:0}}
\end{center}
\end{figure}

\begin{figure}
\begin{center}
\includegraphics[width=\textwidth]{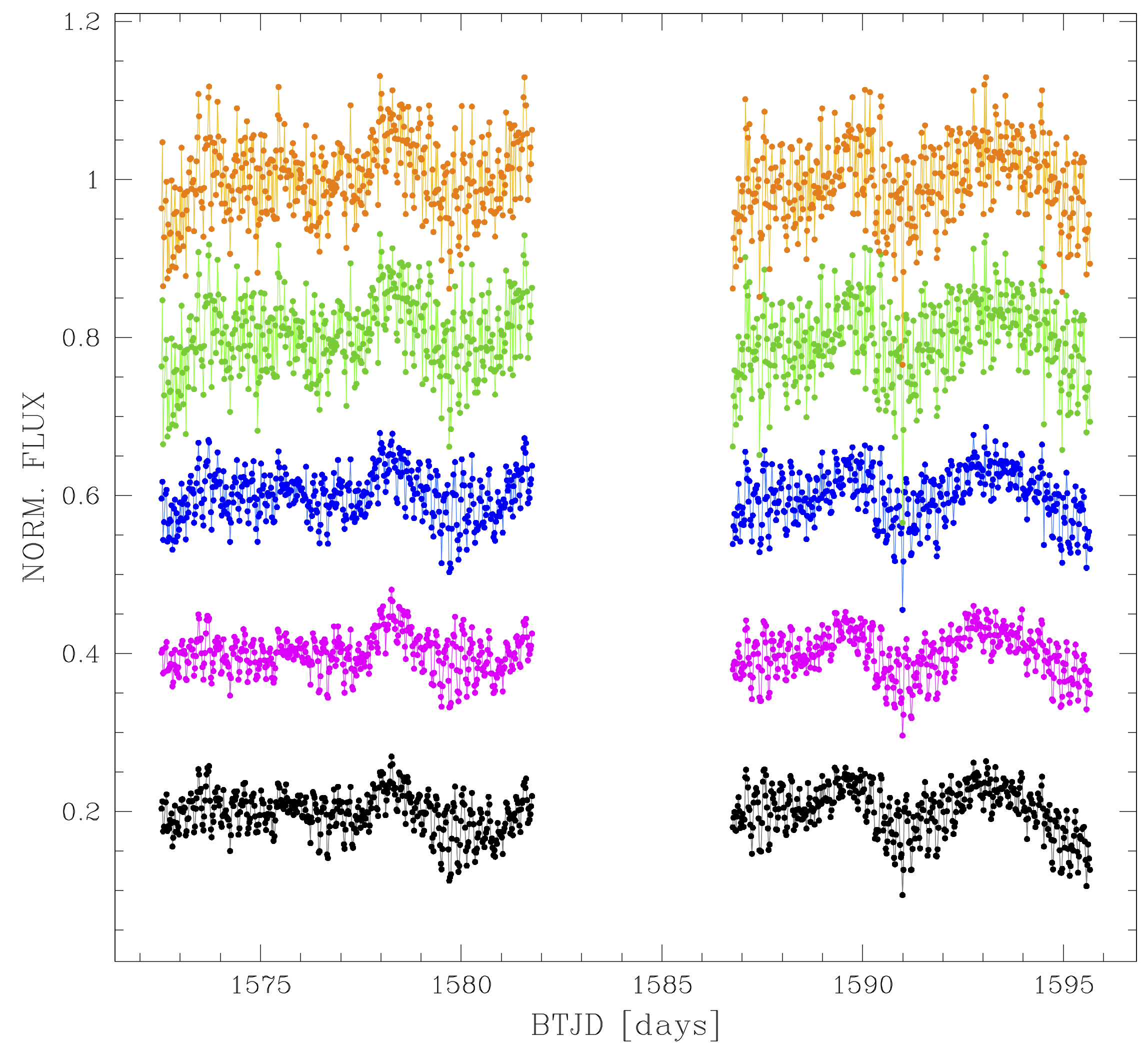}
\caption{ Lightcurves of the Luhman 16AB system extracted from {\it TESS} data obtained with different photometric methods: PSF-fitting (black) and aperture photometry with radius 1-px (magenta), 2-px (azure), 3-px (green), and 4-px (orange).   \label{fig:1}}
\end{center}
\end{figure}

\subsection{TESS Photometry}
\label{S:TESSPhotometry}
Luhman\,16\,AB were observed by the Transiting Exoplanet Survey
Satellite (\textit{TESS}, \citealt{Ricker2014}) during Sector 10,
between March 26th, 2019 and April 22th, 2019. Luhman\,16 system was
only observed on Full Frame Images (FFIs) mode, with a cadence of 30
minutes.

Because of the low resolution, the Luhman\,16AB system is not resolved
on the {\it TESS} FFIs; as a consequence we extracted the combined lightcurve
for the system.

For the extraction of the lightcurve, we used our PSF-based approach described in detail in \citet{Nardiello2019} for the case of 
{\it TESS} time-series, and already adopted for ground- and space-based
time series (see,  \citealt{Nardiello2015,Nardiello2016a,Nardiello2016b} and  
\citealt{Libralato2016a,Libralato2016b,Benatti2019}). 

Briefly, by using on FFIs, accurate PSF models, and a high resolution input catalogue
(in this work the one from \textit{Gaia}\,DR2, \citealt{GAIA2018}), we extracted
high-precision photometry of the target source after subtracting
adjacent field stars using 2 different photometric methods: aperture
photometry (with 1-, 2-, 3-, 4-\textit{TESS} pixel radius) and PSF-fitting photometry.

Because Luhman\,16AB is a close-by high proper-motion source, we calculated
the position of the system at the epoch of observations ($\sim
2019.26$) by using the astrometric positions calculated by \citet{Bedin2017}. 
We adopted $(\alpha_{2019.26},
\delta_{2019.26})=(162.303282427, -53.317573814)$, that corresponds to a time-averaged (over the observing sequence)
detector position with the raw pixel coordinates of $(1767.4,7.4)$ on the CCD 1 of Camera 2 (see Fig.~\ref{fig:0}). 

After the extraction, we corrected the systematic trends that affect
the lightcurve by using cotrending basis vectors (CBVs) extracted
from a set of 2000 stars across the whole CCD. We applied the CBVs as in 
\citet{Nardiello2019}. 

The corrected, normalized lightcurves extracted with all the photometric methods are shown in Fig.~\ref{fig:1}. We computed the \texttt{P2P RMS}, as calculated in \citet{Nardiello2019}, for all the photometric methods, to select the best lightcurve\footnote{\texttt{P2P RMS} is not affected by the variability of the source}. We found that the lower \texttt{P2P RMS} are for the lightcurves obtained with 1-px aperture photometry ($\sim 9.8$\,mmag) and PSF-fitting photometry ($\sim 10.6$\,mmag), while the other photometric apertures show higher \texttt{P2P RMS} ($\sim 14.1$\, mmag, $\sim 28.7$\,mmag, $\sim 42.8$\,mmag, for 2-px, 3-px, and 4-px aperture photometries, respectively).
In this work we use the 1-pixel aperture photometry lightcurve (in magenta in
Fig.~\ref{fig:1}).

\begin{figure}[h]
\begin{center}
\includegraphics[width=1.0\linewidth]{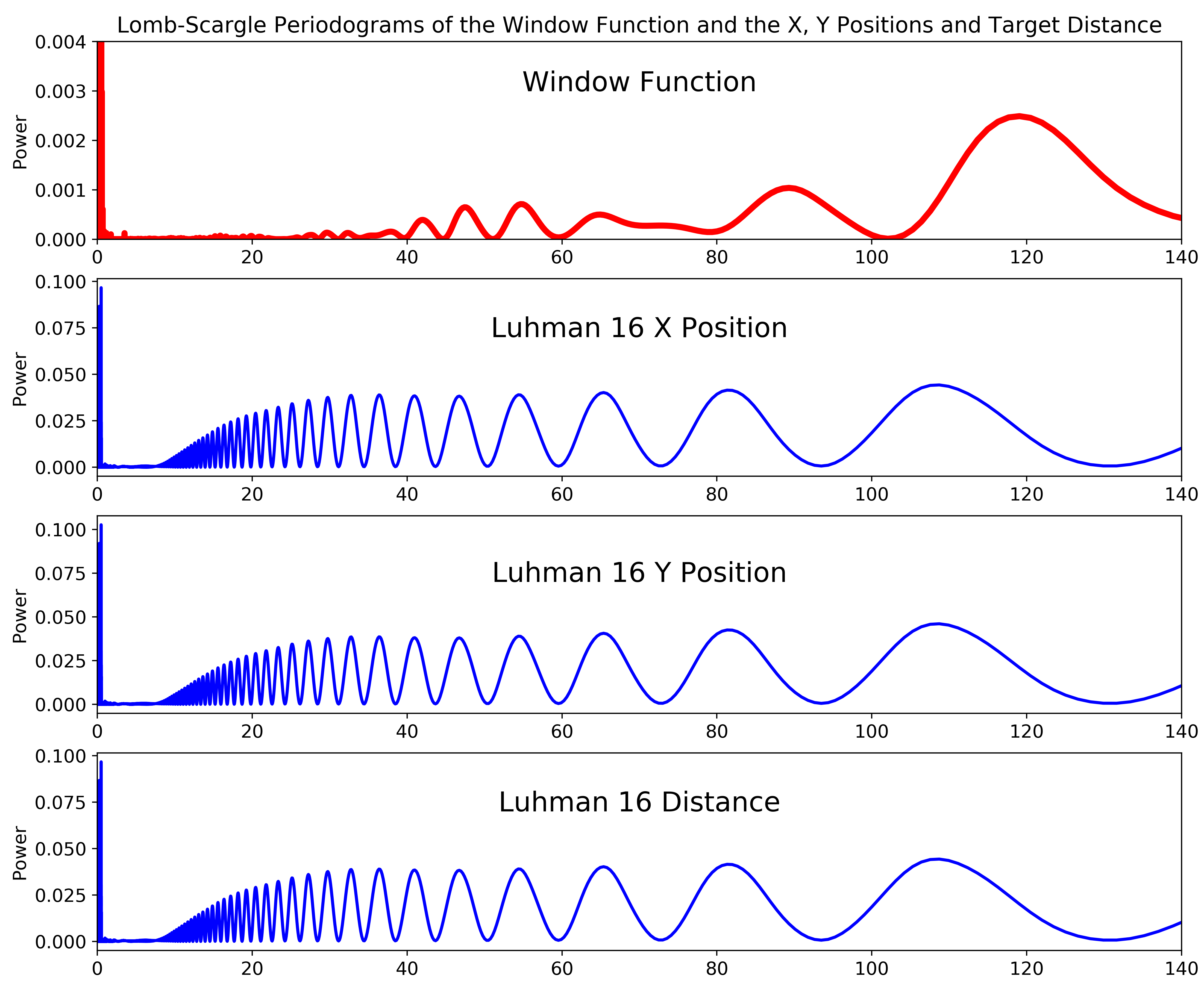}
\caption{
Lomb-Scargle periodograms of the window function (top panel), and the X- and Y-positions of the TESS aperture, and for its distance from the median (X,Y) coordinate. The LS periodograms of the spacecraft position have multiple peaks in the explored period range. However, the short-period (P$<$20~h) range is clean of major peaks as is the vicinity of the long-period (P$\sim$90.8~h) peak identified in the lightcurve. Therefore, we conclude that slight spacecraft positional jitter does not explain or expected to impact the periods identified as physical in the lightcurve.\label{fig:PositionalLS}}
\end{center}
\end{figure}

{In much of the current study we explore the periods present in the TESS lightcurve of \luh{}AB. Given the relatively large pixels of TESS, false periods appearing due to spacecraft positional jitter may be a concern. In other words, a brighter star within or close to the aperture may contribute to the measured light and, if the telescope's position is oscillating, it may introduce an apparent periodicity. Given TESS's extremely stable pointing such a contamination is very unlikely. Nevertheless, in order to conservatively test this possibility, we calculated the Lomb-Scargle periodogram \citep{Lomb1982,Scargle1976} the same way as we do for the lightcurve analysis in subsequent sections (Section~\ref{S:Periodogram}). The resulting periodogram is shown in Figure~\ref{fig:PositionalLS}, along with the window function (top panel). We include the L-S periodograms of the x-- and y--positional differences, as well as the difference from the aperture center. The Lomb-Scargle periodogram does display peaks, demonstrating a very slight oscillation with multiple periods. However, comparison of these periodograms to the peaks in the periodogram of the astrophysical source shows that none of the peaks identified in \luh{}AB's lightcurve correspond to peaks in the periodogram of the aperture position. The lack of match between the peaks in the \luh{AB and the }positional modulation periodograms demonstrates that the \luh{}AB lightcurve periodicity \textit{does not} emerge from spacecraft oscillations.  }

The lightcurve will be released on the MAST archive as HLSP under the project 
PATHOS\footnote{\url{https://archive.stsci.edu/hlsp/pathos}, doi:10.17909/t9-es7m-vw14} (see \citealt{Nardiello2019} for details).

\subsection{HST Lightcurves}
\label{S:BackgroundStars}
{Another potential, though unlikely source of contamination in our TESS \luh{}AB lightcurve could be introduced by background stars within or close to the TESS aperture. Any star that would introduce such a contamination would need to be relatively bright (comparable to Luhman 16AB) and would need to display strong variability (at multi-percent level). Although TESS does not allow for easy disambiguation of multiple sources in the aperture, we were able to utilize already published Hubble Space Telescope Wide Field Camera 3 Ultraviolet/Visible channel (HST/WFC3/UVIS) image series, obtained in the broad-band F814W filter. The image mosaic covers an area of about 160$^{\prime\prime}\times160^{\prime\prime}$. These images have been collected between 2014 and 2016 in 12 epochs, with the goal to monitor the astrometric evolution of the binary system. They have been analyzed and published in \citet[][]{Bedin2017}. These datasets included high-precision photometric measurements for 808 stars in the vicinity of \luh{}AB. }

\begin{figure}[h]
\begin{center}
\includegraphics[width=1.0\linewidth]{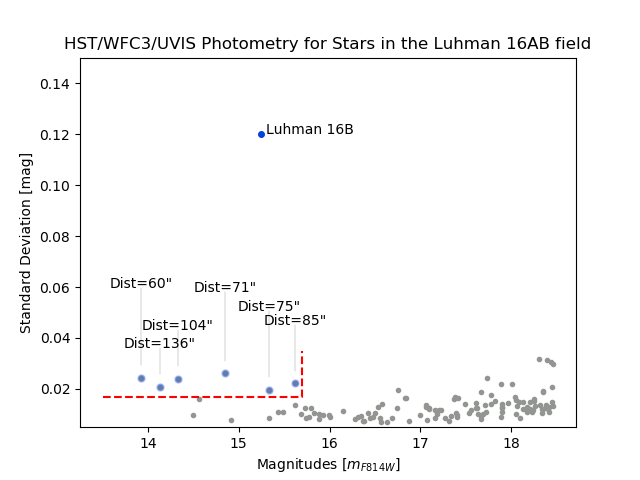}
\caption{Brightness and standard deviation of the stars in the Luhman 16AB field from HST photometry (12 epochs, \citealt[][]{Bedin2017}). All stars that are comparable in brightness to \luh{}AB and show some variability (blue symbols) are far in projection from \luh{}AB and very unlikely to contaminate the TESS aperture. In addition, none of the stars shows variability approaching that detected in the TESS lightcurve of \luh{}AB. We conclude that contamination from a field star is extremely unlikely.  \label{fig:HST-LS-Scatter}}
\end{center}
\end{figure}


{Our goal was to test the hypothesis that one or more relatively bright star exists within the vicinity of the Luhman16AB aperture. We thus searched the multi-epoch photometry for bright, high-amplitude variable stars close to \luh{}AB. In Figure~\ref{fig:HST-LS-Scatter} we plot the measured brightness (F814W band) of stars against the standard deviation of their lightcurves (as a a measure of their potential variability). \added{For reference, we also plot \luh{}B's observed brightness and semi-amplitude from \cite[][]{Bedin2017}.}
In the figure we highlighted the six stars that are of comparable brightness to \luh{}AB ($m_{\rm{814W}}\sim$15.2~mag, \citealt[][]{Bedin2017}). For each of these stars we  also include labels that show the projected distances of these stars from \luh{}AB. Based on this information, we conclude that there are no stars of comparable brightness to \luh{}AB within the TESS aperture; and that the closest, relatively bright star is 60\arcsec{} away, i.e., about 3$\times$ the radius of the TESS aperture used in our study. These facts make it unlikely that any of the stars contaminate the lightcurve. Furthermore, as explained in the following sections, the \luh{}AB TESS lightcurves display large-amplitude variations (up to $\sim$10\%); this level of variation is about $4\times$ higher than the most variable stars seen in our data. We note a minor caveat to this analysis: the central pixels of the few brightest stars may be non-linear in the HST images, thus, their brightness may be slightly underestimated \citep[][]{Bedin2017}. Nevertheless, the rest of the pixels are well sampled and the low standard deviations further bolster confidence in our findings.
 As one final step and additional test, we also phase-folded the lightcurves of the ten stars with the 90.8~h period, a particular period identified later in this analysis. This analysis also turned up negative: none of the stars showed even low-amplitude variability with the 90.8~h period. }
 
{In summary, we conclude that due to the lack of bright stars in the vicinity of \luh{}AB and due to the lack of highly variable stars anywhere in the field, it is extremely unlikely that the periods observed in the \luh{}AB TESS lightcurves are significantly contaminated by a background star.}

%
\section{Results: Modulations and Lightcurve Evolution}
During the entire duration of the dataset, the LC of \luh{} remained variable. The greatest relative brightness difference observed in the dataset was about $\sim$13\% (between the maximum and minimum of Segment 1), but the typical peak-to-trough brightness changes \textit{over the time-scale of single rotation} were typically about 4\%. The LC shows long-term modulations, as well as short-term ($\sim$ 1 rotational period) modulations. The \luh{} lightcurve displays a complex, yet familiar behavior (see Figure~\ref{fig:Overview}), reminiscent to those reported in short-duration datasets for Luhman 16AB (\citealt{Gillon2013,Buenzli2015a,Buenzli2015b,Karalidi2016}) and other L/T brown dwarfs (\citealt{Apai2017}).  The lightcurve shows dozens of peaks over the nearly 50-day long baseline of the observations, but no obvious, simple periodicity; instead, continuous evolution is apparent in the lightcurve.

Based on the inspection of the lightcurve we conclude the following: (1) \luh{} remains variable during the entire data set, covering over 540~h, suggesting that the atmosphere itself remained heterogeneous. (2) The lightcurve is not strictly periodic but continues to evolve, confirming previous reports of rotational modulations and lightcurve evolution \citep{Gillon2013,Karalidi2016} in \luh{}. (3) Both short-term (comparable to the rotational period of $\sim$5.4\,h, see Discussion) and long-term (60--100~h) variations are apparent. 

{In the following we will explore the lightcurve evolution via a Lomb-Scargle periodogram analysis (Section~\ref{Section:PowerSpectrumAnalysis}), followed by an exploratory lightcurve modeling with the planetary-scale wave model (Section~\ref{Section:LCAnalysis}).}

%
\section{Periodogram Analysis}
\label{S:Periodogram}

We now turn to a Generalized Lomb-Scargle periodogram analysis (see \citealt[][]{Scargle1976,Lomb1982,ZechmeisterKurster2009}, and, for a review of applications and limitations, \citealt{VanderPlas2018}) to explore the underlying periodic components within the \luh{} intensity variations. This section presents the periodogram analysis, along with the study of potential window function artifacts, and the results of the analysis. 

\subsection{Lomb-Scargle Periodogram and Window Function Analysis}

Lomb-Scargle periodogram analysis can be performed on data sampled on an irregular grid, but such datasets will inevitably introduce a non-uniform power distribution in the periodogram. Although the TESS dataset provides an excellent lightcurve with close to perfect sampling, the impact of the window function on the periodogram must always be carefully considered. 
To aid the interpretation of the Lomb-Scargle periodogram we calculated the window function's Generalized Lomb-Scargle periodogram (using the PyAstronomy package, \citealt[][]{Czesla2019}, with Zechmeister--K\"urster normalization, \citealt{ZechmeisterKurster2009}). The window function was calculated over a finely and regularly sampled array where the function's value was set to unity at point when measurements were available and zero elsewhere. This window function was then analyzed through the Lomb-Scargle method to allow the identification of the sampling's imprint on the periodogram.

\begin{figure}
\begin{center}
\includegraphics[width=\textwidth]{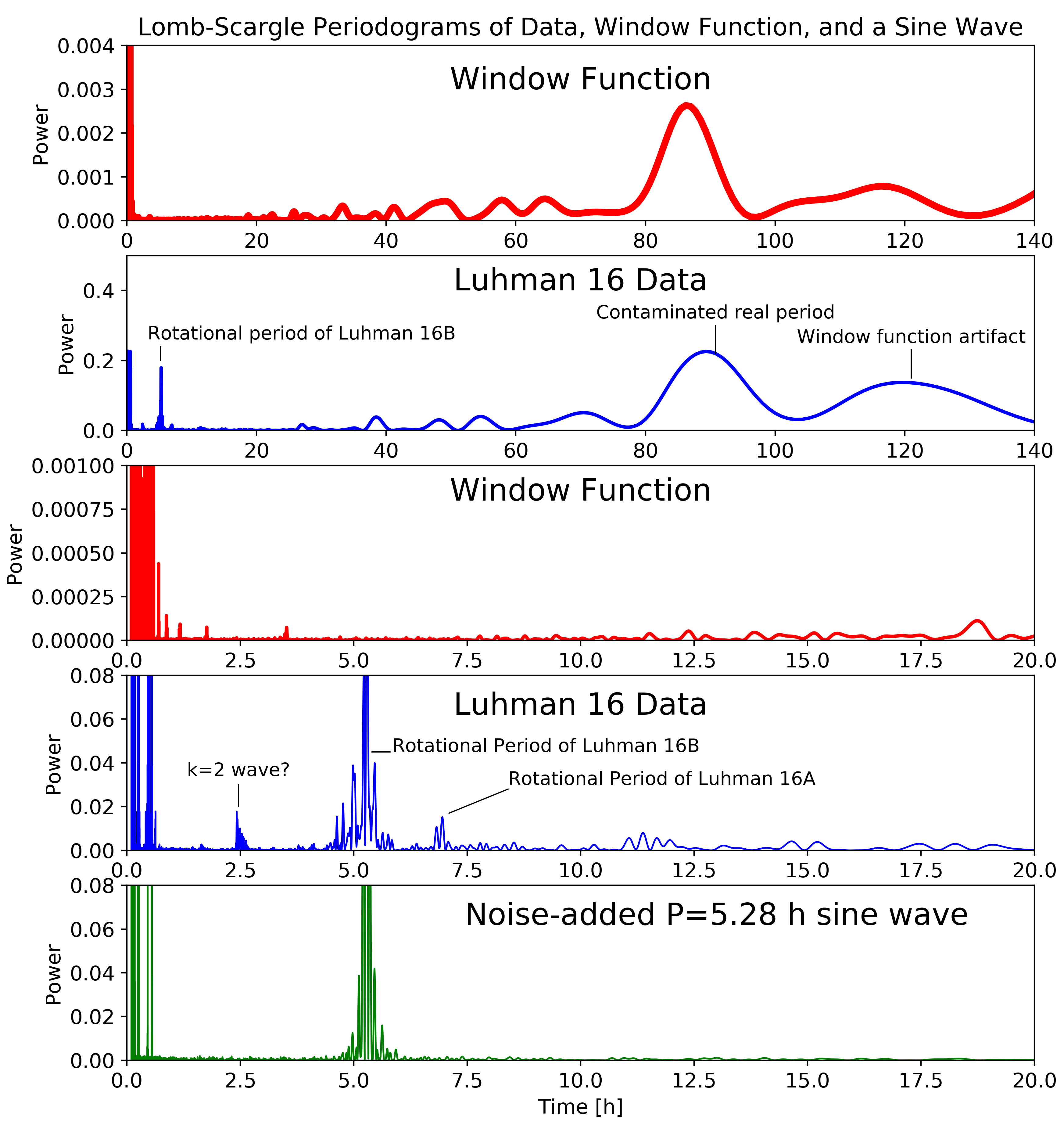}
\caption{Lomb-Scargle periodograms of the TESS data (blue), its window function (red), and a noise-added sine wave (green) with the same temporal sampling as the data. The top four panels show the same data and window periodograms but for different period ranges, to allow closer inspection. The 1--20~h period range is not affected by the window function. A clear physical periodicity is identified at P=5.28~h. Additional possible periods may be present at P=2.5~h and P=6.94~h.
\label{fig:LS-Window}}
\end{center}
\end{figure}

In Figure~\ref{fig:LS-Window} we show the resulting Lomb-Scargle periodograms for the Luhman~16 dataset (in blue) and for the window function (red). The two upper panels show the 0--140~h period range, while the two  panels below them show the same for the 0--20~h period range, allowing a finer inspection. As comparison, the bottom panel shows the periodogram of a noise-added sine wave function sampled at the same time as the TESS observations. The period of the sine wave is P=5.28~h and its amplitude is 0.05. The noise added to this sine wave resembles the photon noise estimated for our data (1$\sigma$=0.013).

Based on Figure~\ref{fig:LS-Window} we conclude that the window function's periodogram is nearly featureless for periods between 1--20~hours, due to TESS's efficient sampling of the lightcurve. The window function, however, displays prominent broad and prominent peaks at 86~h and 120~h, and additional multiple narrower and weaker peaks corresponding to periods between 20 and 64~h. 

\subsection{Interpretation of the Peaks}
Our periodogram analysis leads to the following conclusions:

(1) A confident detection of periodicity corresponding to 5.28~h, which likely corresponds to the rotational period of the \luh{} component that dominates the rotational modulations. 
(2) The peak corresponding to the P=5.28~h period modulations is slightly broader in the data than that of the sine wave (bottom panel).  {We explore this finding in more details in Section~\ref{s:Peaks}.}

(3) Two additional, but much weaker peaks are observed at P=2.5~h and P=6.94~h. These do not correspond to peaks that would emerge from the window function or from the P=5.28~h sinusoidal modulation.  Peaks in power spectra corresponding to half-periods have been reported in Spitzer lightcurves of brown dwarfs \citep[][]{Apai2017}, and this may explain the P=2.5~h period peak. The P=6.94~h period peak, we speculate, may correspond to the rotational period of the \luh{} component that has lower amplitude in the TESS-band. We note that through injection of sine waves with these periods we verified that these two peaks are {\em present independently} of each other, i.e., neither the 2.5~h or the 6.94~h peak is an alias of the other peak, or emerges from a combination of the other peak and the 5.28~h peak. 

(4) We also identify two peaks in the power spectrum that correspond to long-term modulations: A peak at period 90.81~h and a peak at period 126.6. Both of these peaks lie very close to peaks prominent in the window function's periodogram, which raises the possibility that these are not physical features. To help their interpretation, we calculated the false alarm probabilities (FAPs) for these peaks. We found that the P=5.28~h and P=90.81~h peaks have very low false alarm probabilities (log(FAP) of $-46$ and $-44$, respectively). In contrast, the P=126.6~h peak's log(FAP) is much higher ($-13$). Given these considerations, we conclude that the peak at period 90.8~h is likely physical, while the peak at period 126.6~h is likely not physical. Although the window function complicates the interpretation of such long-period peaks, supporting our interpretation is the fact that the TESS lightcurve shows very clear long-term evolution (see, for example, Figures~\ref{fig:1} and \ref{fig:Overview}). Although a period close to 90.8~h is very likely real, given the contamination from the window function, we consider its exact period not well determined.


\subsection{Fine Structures of the 5.28~h and 2.50~h Peaks}
\label{s:Peaks}

\begin{figure}
\begin{center}
\includegraphics[width=\textwidth]{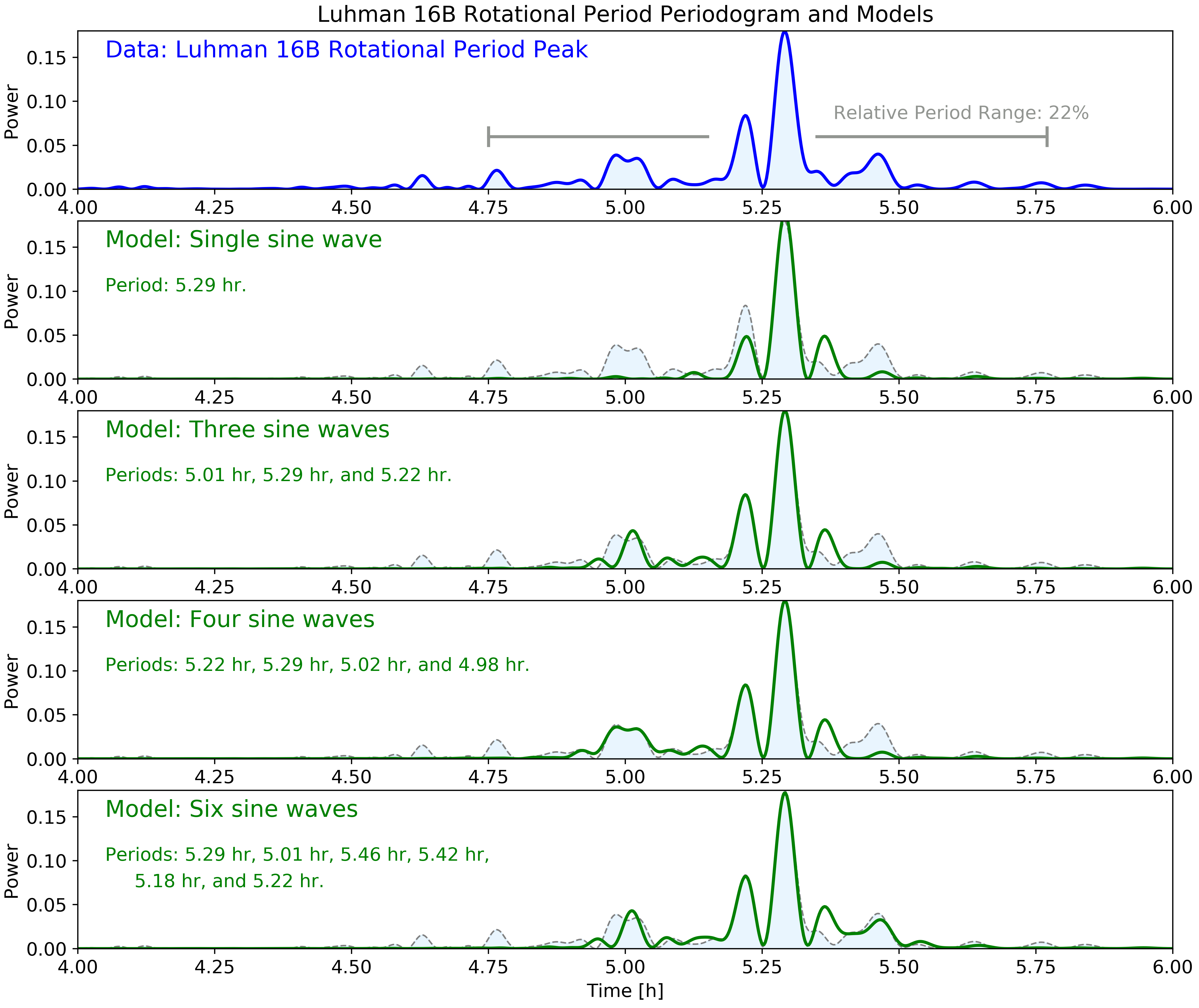}
\caption{Lomb-Scargle periodograms of the TESS data (blue), focusing on the rotational period of \luh B, and models consisting of sine waves modeled in the period space.
\label{fig:MultiplePeaks}}
\end{center}
\end{figure}

 {We will now explore the nature of the P=5.28~h peak in the periodogram. In Figure~\ref{fig:MultiplePeaks} we plot the periodogram data (in blue, top panel) for the P=4.0--6.00~h period range), along with simple model-fits (panels below top panel, in green). The periodogram data (blue) displays multiple peaks: cursory inspection suggests possible peaks around 4.75~h, 5.00~h, 5.22~h, 5.28~h, and 5.45~h. Of course, the periodogram peaks are not trivial to interpret: a single sinusoidal modulation will introduce multiple peaks in the periodogram (see, for example, bottom panel of Figure~\ref{fig:LS-Window} or panel b of Figure~\ref{fig:MultiplePeaks}). We attempt to interpret the observed complex periodogram signature by models of increasing complexity consisting of periodogram signatures of sine waves. Our models are a single sine wave (panel b), three sine waves (panel c), four sine waves (panel d), and six sine waves (panel e). The model fits were carried out by minimizing the sum of the squares of the data$-$model residuals through a Levenberg-Marquardt optimization. Our models are linear combinations of the periodograms of a sine wave, shifted in period-space and scaled in intensity (corresponding to shifts in period and intensity). In order to ensure that the ML-optimization is not trapped in a local minima we repeated the optimization several thousand times, starting from randomly-determined initial guesses, and accepted the best-fit model. The similarities of the solutions for different model complexities demonstrate that our modeling procedure is very robust. }

 {We find that a single sine wave fails to match the complex periodogram signature, i.e., there is clear evidence for the presence of multiple periods. We find that models of increasing complexity succeed in reproducing well the multi-peaked structure of the periodogram (panels c--e in Figure~\ref{fig:MultiplePeaks}). Our most complex model has six periods (panel e), which fits the periodogram of the data very well, although not perfectly. Our model fits provide compelling evidence for the presence of multiple, similar periods in Luhman~16B. }

 {Our fits show the following period (in order of decreasing power): 5.29~h, 5.22~h, 4.98~h, 5.46~h, and 5.02~h. We stopped at a model with six sine waves and did not increase further the model's complexity. We opted to do this because,  while more complex sine-wave based fits are possible, the high-quality data on the fine structure of the periodogram calls for future models more deeply rooted in physical models, such as comparison to predictions by global circulation models \citep[][]{Showman2019}, the predictions of which may be converted to periodograms and compared to the observations.}

 {As a first-order metric to describe the width of the substructure, we provide an estimate for the relative range of periods contained in it. Expressed as ${(P_{max}-P_{min}) \over P_{min}}$ and adopting $P_{min}=4.75$~h $P_{max}=5.77$~h values, we find that the relative period range is $\sim$22\%. }

\begin{figure}
\begin{center}
\includegraphics[width=\textwidth]{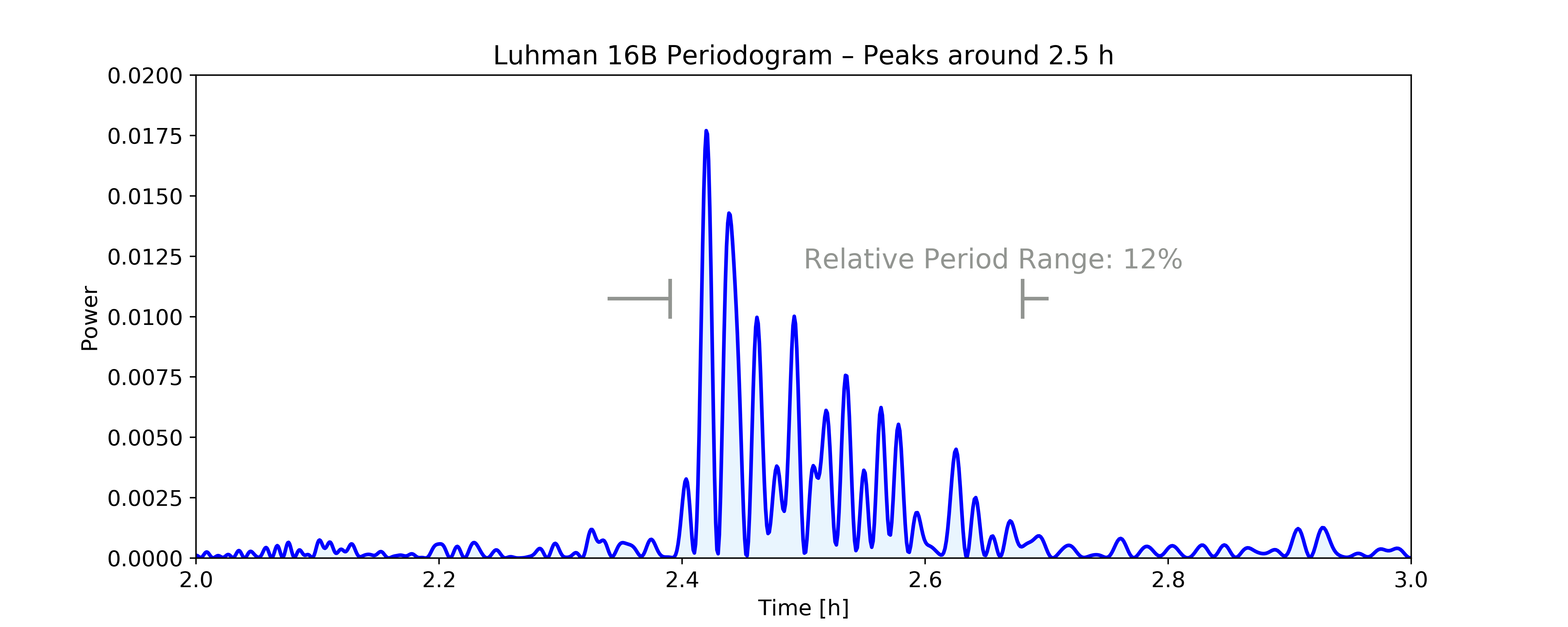}
\caption{Lomb-Scargle periodograms of the TESS data on Luhman~16B, focusing on the peaks around period P=2.5~h. \label{fig:MultiplePeaks2.5}}
\end{center}
\end{figure}

 {The planetary-scale wave model by \citet[][]{Apai2017} would predict that, in addition to k=1 wavenumber waves, k=2 waves would be present. It is thus instructional to explore whether the P=2.5~h peak also shows a multi-peaked substructure as the P=5.28~h peak does. In Figure~\ref{fig:MultiplePeaks2.5} we magnified the Lomb-Scargle periodogram, focusing on the period range 2.0 to 3.0~h. Indeed, the periodogram displays a substructure here, too, with multiple peaks in the approximate period range 2.4 to 2.7~h. We note that as there is much less power in this peak than in the 5.28~h peak, the signal-to-noise ratio is more limited, and the substructure is less well-defined. Therefore, we will not attempt here the detailed modeling of the substructure. Instead, to allow a first-order comparison to the 5.28~h peak}
structure, we calculate the approximate relative period range in which peaks appear to above the background noise. Expressed as ${(P_{max}-P_{min}) \over P_{min}}$ and adopting $P_{min}=2.4$~h $P_{max}=2.7$~h values, we find that the relative period range is $\sim$12\%. 

 {In summary, we conclude that both the P=5.28~h and the P=2.5~h peaks display fine structures, with multiple peaks; and that the peaks in the 2.5~h group are likely the k=2 (half-period) waves corresponding to the k=1 waves in the 5.28~h group. The high-quality P=5.28~h peak group reveals the presence of multiple -- at least six -- distinct, slightly different rotational periods. }

\subsection{Long-period Modulation (P=90.8~h)}

{We note here, however, that there is a remote possibility that the long-period modulation (P=90.8~h) is introduced by a background star that falls within the same aperture. Although we have done extensive testing based on the available data and found no indication for such a scenario, this cannot be excluded with the present data at hand. In the future TESS campaigns, however, by re-visiting the high proper-motion \luh{}, contamination by a background star (with negligible proper motion) will be possible to identify.}

\label{Section:PowerSpectrumAnalysis}

%
\section{Lightcurve Modeling}
\label{Section:LCAnalysis}

\begin{figure}[h]
\begin{center}
\includegraphics[width=1.0\linewidth]{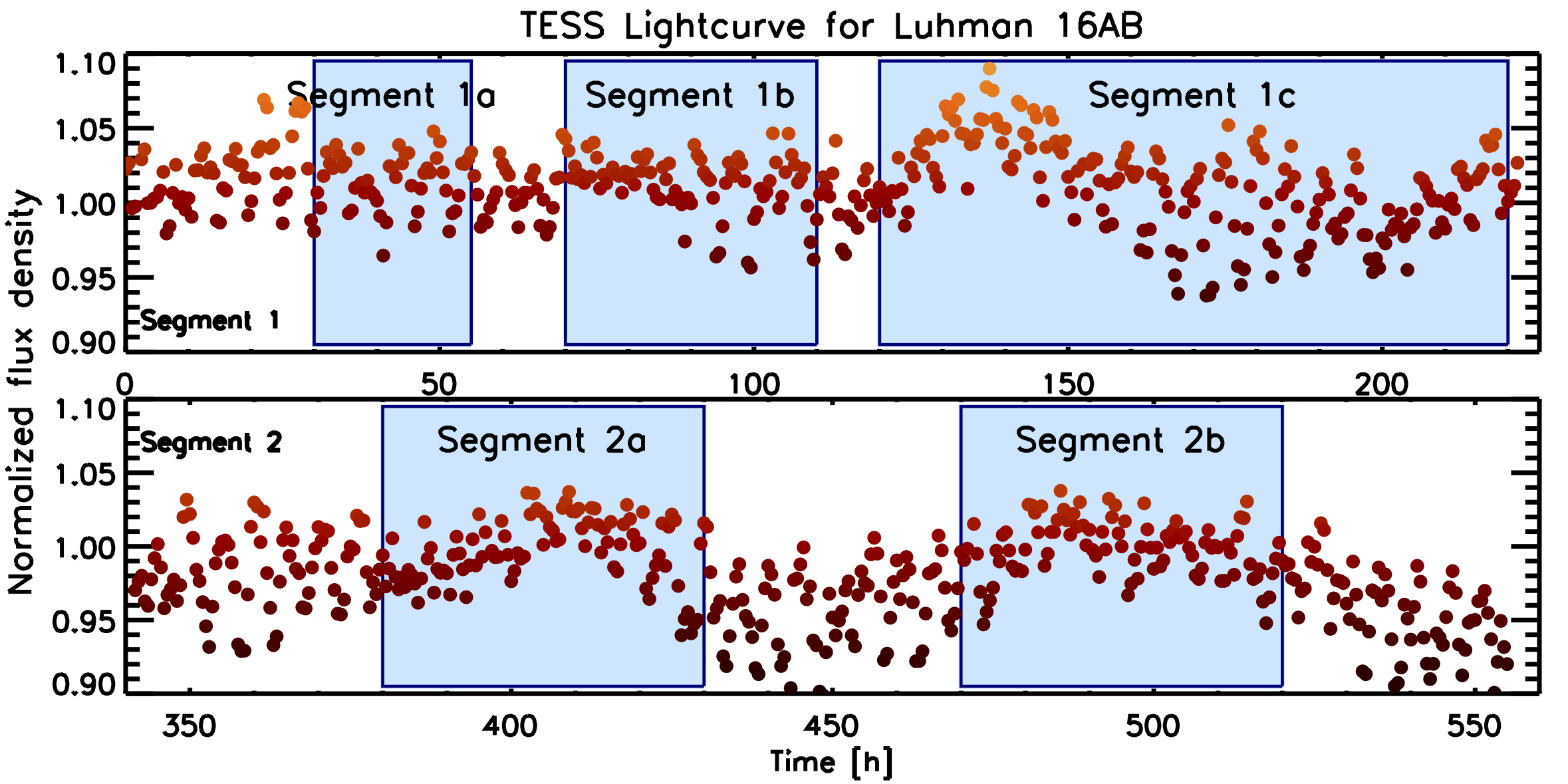}
\caption{Overview of the TESS lightcurve of Luhman 16~AB. Within the two segments
four regions are marked (Segments 1abc, 2ab), which are magnified and modeled in Figures~\ref{fig:Segment1a},~\ref{fig:Segment1b}, \ref{fig:Segment1c},\,\ref{fig:Segment2a}, and \ref{fig:Segment2b}.
\label{fig:Overview}}
\end{center}
\end{figure}

\begin{figure}[h]
\begin{center}
\includegraphics[width=1.0\linewidth]{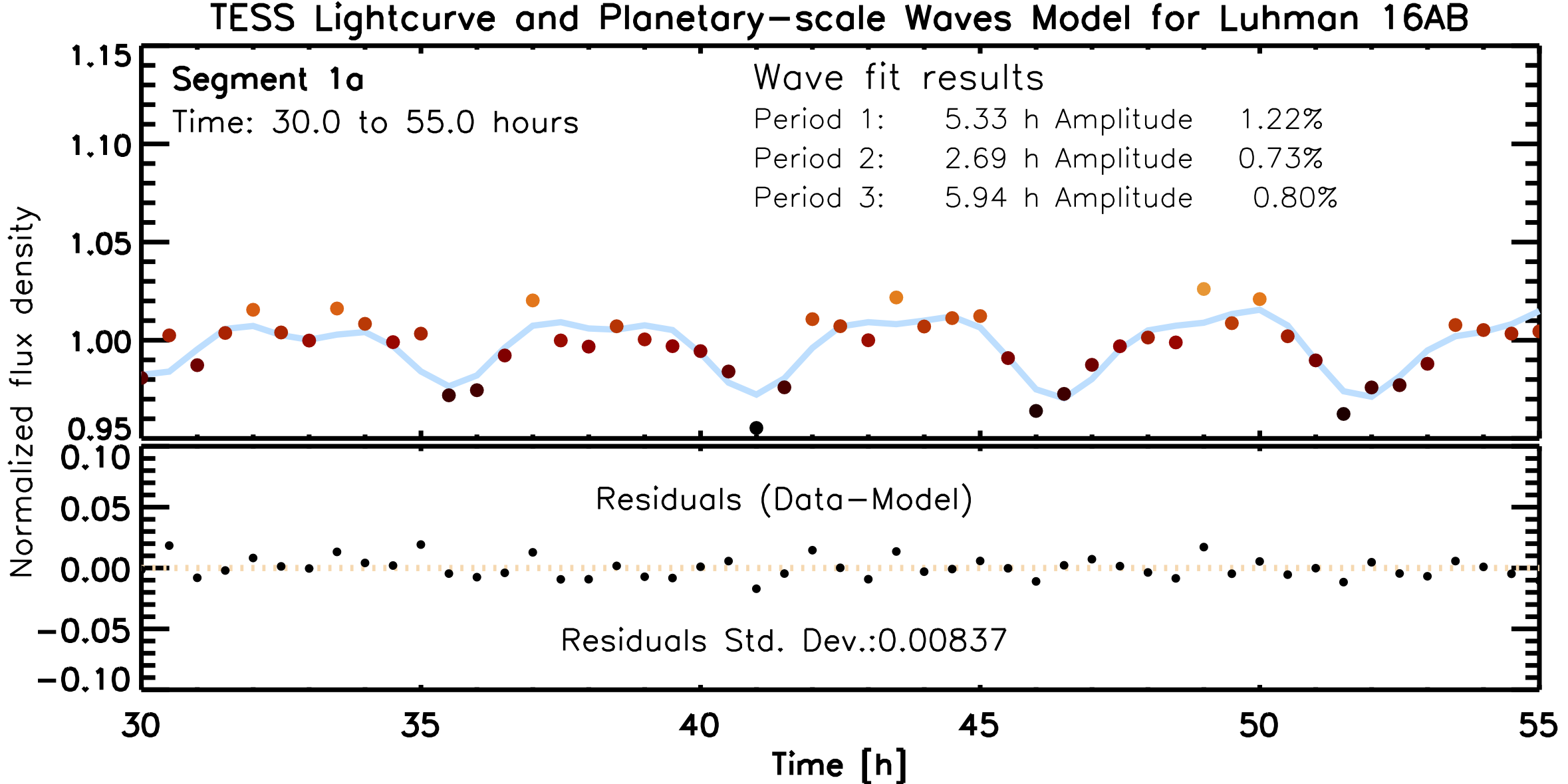}
\caption{\textit{Upper panels:} Segment 1a of the  \luh{} lightcurve  {(minus long-term trend)} and the planetary-scale waves model (blue curve). \textit{Bottom panels:} Residuals (data-model) and the model. The planetary-scale waves model  {reproduces well the evolution of the lightcurve with fitted period values fully consistent with those found in the Lomb-Scargle analysis.} \label{fig:Segment1a}}
\end{center}
\end{figure}

\begin{figure}[h]
\begin{center}
\includegraphics[width=1.0\linewidth]{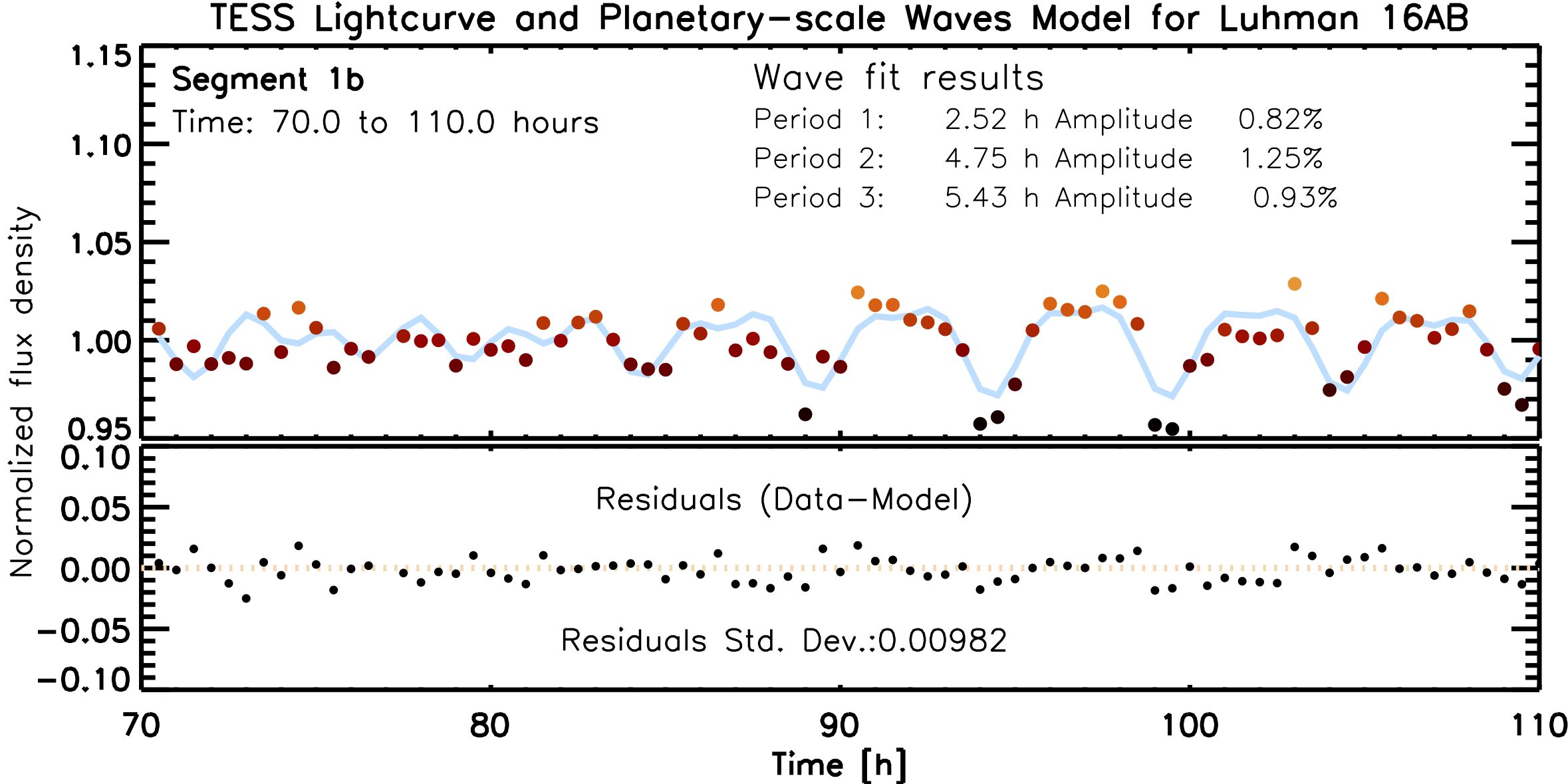}
\caption{
\textit{Upper panels:} Segment 1b of the \luh{} lightcurve  {(minus the long-term trend)} and the planetary-scale waves model (blue curve). \textit{Bottom panels:} Residuals (data-model) and the model.  {reproduces well the evolution of the lightcurve with fitted period values fully consistent with those found in the Lomb-Scargle analysis.}
\label{fig:Segment1b}}
\end{center}
\end{figure}

\begin{figure}[h]
\begin{center}
\includegraphics[width=1.0\linewidth]{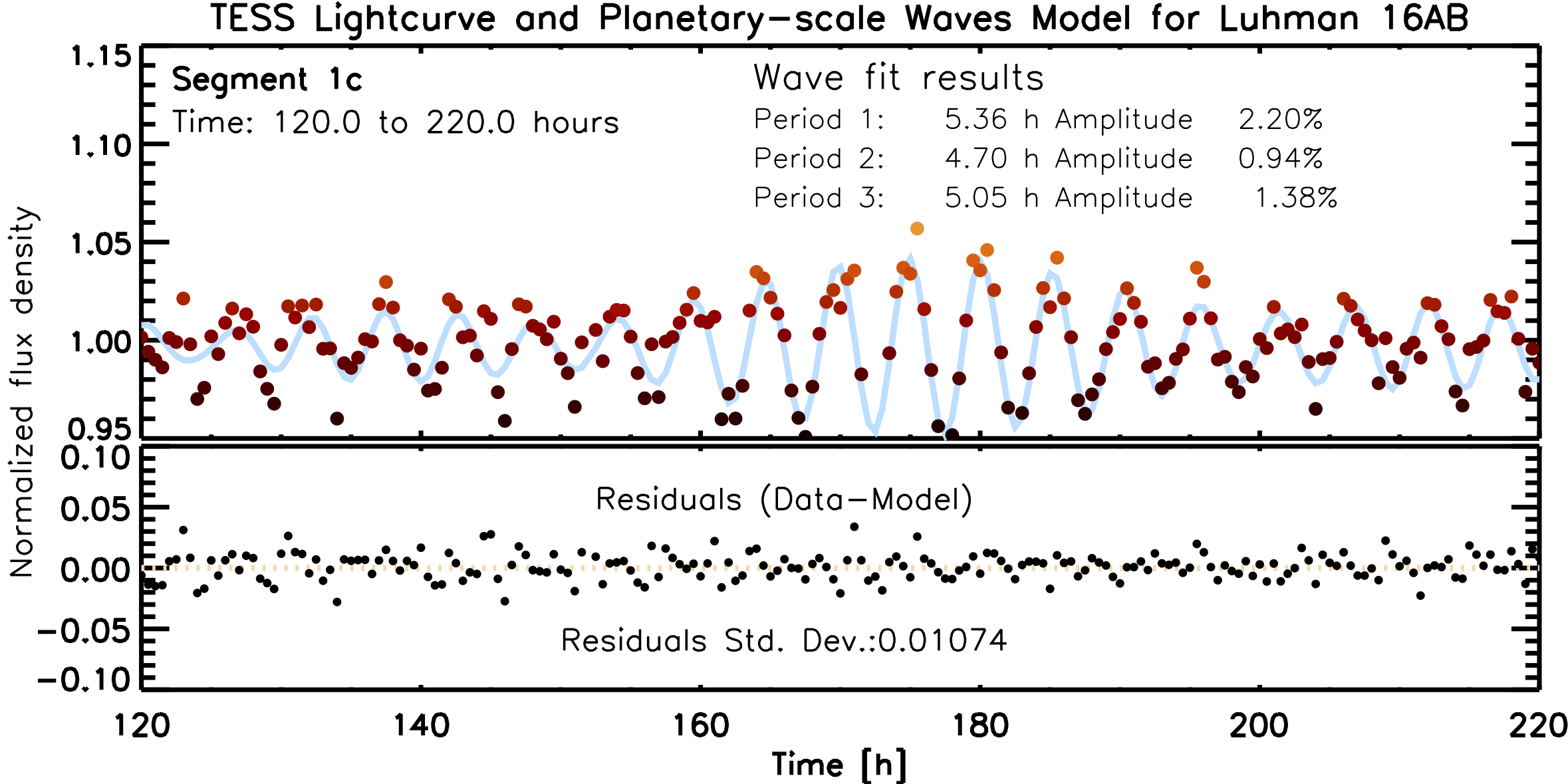}
\caption{\textit{Upper panels:} Segment 1c (100~h)of the \luh{} lightcurve  {(minus long-term trend)} and the planetary-scale waves model (blue curve). \textit{Lower panels:} Residuals (data-model) and the model. Covering about 20 rotational periods, our longest segment   {modeled is still reproduced well by the planetary-scale waves model, with fitted period values fully consistent with those found in the Lomb-Scargle analysis.} \label{fig:Segment1c}}
\end{center}
\end{figure}

\begin{figure}[h]
\begin{center}
\includegraphics[width=1.0\linewidth]{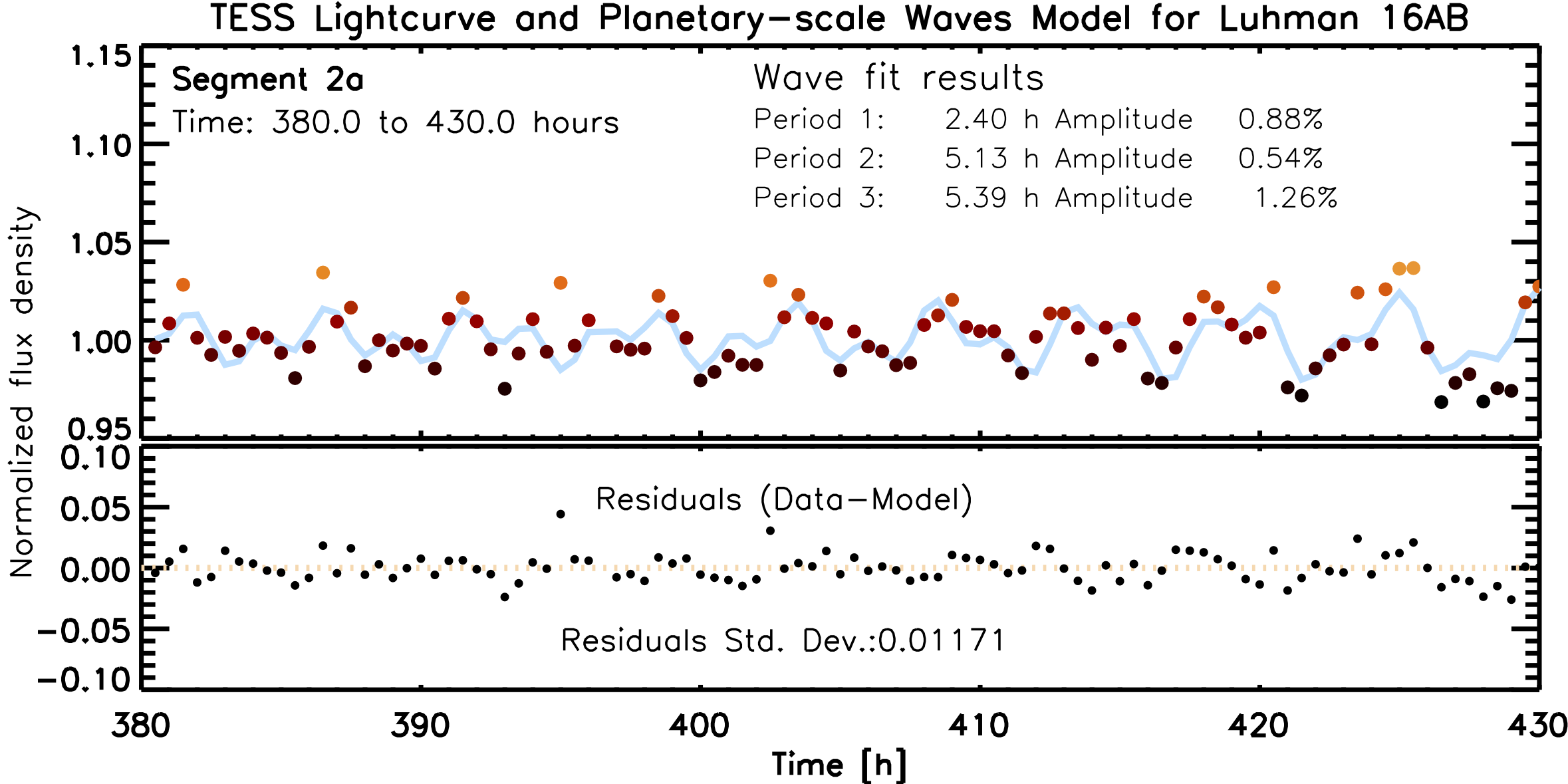}
\caption{\textit{Upper panels:} Segment 2a of the \luh{} lightcurve  {(after subtraction of the long-term trend)} and the planetary-scale waves model (blue curve). \textit{Bottom panels:} Residuals (data-model) and the model. \label{fig:Segment2a}}
\end{center}
\end{figure}

\begin{figure}[h]
\begin{center}
\includegraphics[width=1.0\linewidth]{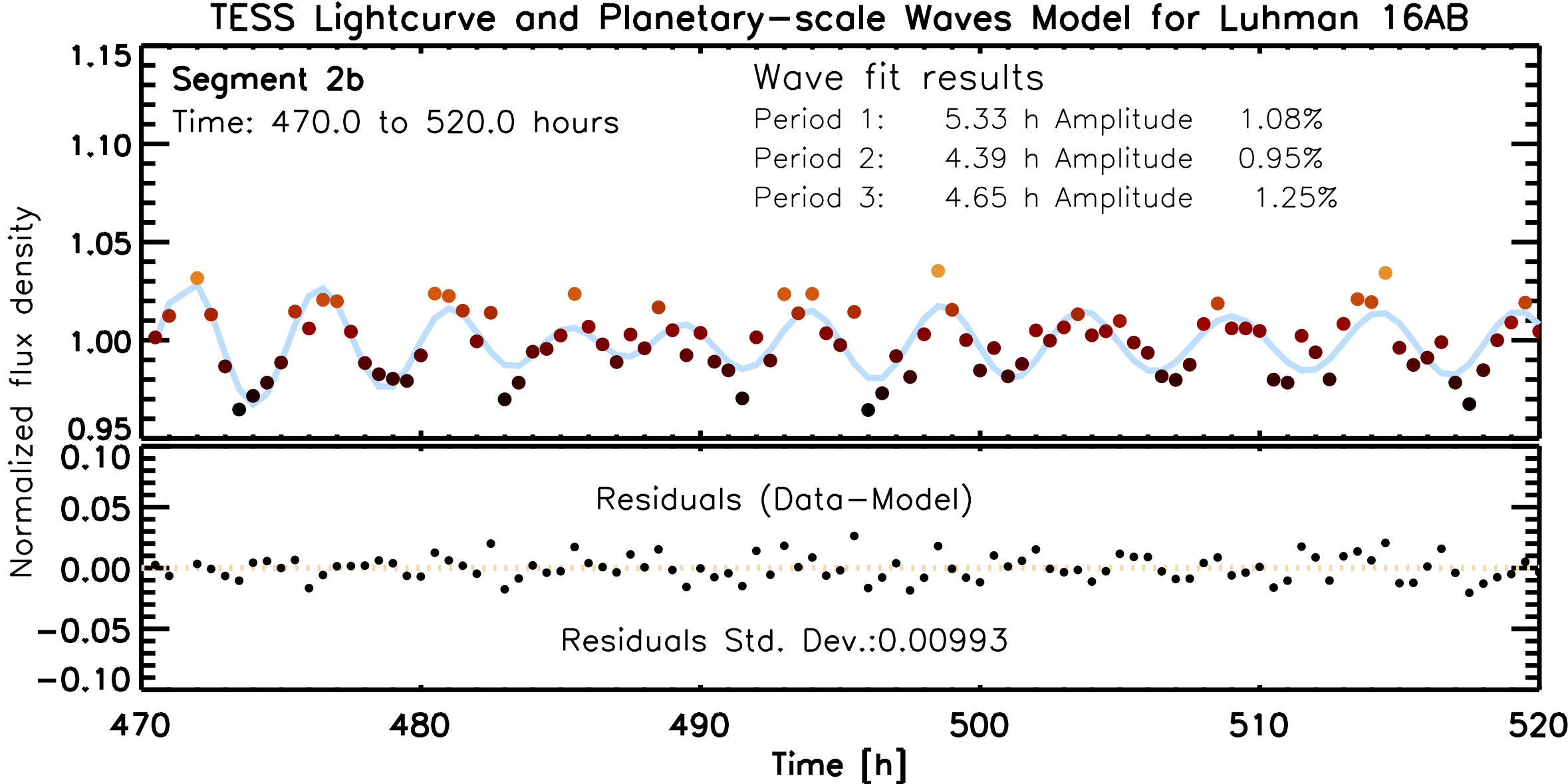}
\caption{\textit{Upper panels:} Segment 2b of the \luh{} lightcurve  {(after subtraction of the long-term trend)} and the planetary-scale waves model (blue curve). \textit{Lower panels:} Residuals (data-model) and the model. \label{fig:Segment2b}}
\end{center}
\end{figure}

 {Our Lomb-Scargle analysis showed that in Luhman 16B's lightcurve multiple peaks are present with slightly different periods, in two groups: one centered on the rotational period of the object $\sim$5.28~h and one centered at the half-period $\sim$2.52~h. These findings are fully consistent with the lightcurve evolution model proposed by \citet[][]{Apai2017}.}
That model is based on the comprehensive Spitzer/IRAC monitoring obtained in \citet{Apai2017} and explains the lightcurve evolution with planetary-scale waves with similar amplitudes, but slightly different periods (likely due to differential rotation). \citet{Apai2017} also reported the presence of k=1 and k=2 waves (where k=1 corresponds to a single peak during a rotational period). 
 {We will now explore how well this model can actually fit the TESS lightcurves of Luhman~16\,AB presented here.}

We applied the same model described in \citet{Apai2017} to the TESS lightcurve with a few minor modifications. These changes are required to run the algorithm efficiently on the TESS LC, but are not expected to lead to any significant difference between the model published in \citet{Apai2017} and our implementation.  For the details of the model and extensive test results we refer to \citet{Apai2017} and its supplementary online material, and here only summarize the fundamental setup of the model and the minor changes unique to our implementation for the TESS data.
 
The planetary-scale wave model describes the evolving lightcurve evolution as a sum of three sine waves. The periods, amplitudes, and phases of the three sine waves are mostly unconstrained. The three sine waves are fit through a two-stage fitting process: first a non-heuristic algorithm (Genetic Algorithm) is used to identify the likely global optimum, which is then passed on as initial guess to an efficient multi-dimensional gradient optimizer to find the exact best-fit solution. 

 {We made two modifications to the original procedure: First, we constrained the range of the periods fitted to 2--6 hours, the full range of periods identified by the Lomb-Scargle analysis. This ensured that the parameter space searched is consistent  with the periodogram results, but without prescribing or favoring actual periods within that range. Second, in our model we did not attempt to fit the very long-period (P=90.8~h) component of the lightcurve, but focused on the evolution of the short-period component. Therefore, we fitted a copy of the lightcurves with the long-term evolution removed by subtracting a 10-sample (5~h window) boxcar-smoothed version of itself \added{(fits using a 7.5~h boxcar window led to identical results)}. }

Figure~\ref{fig:Overview} shows the entire lightcurve split into two segments (Segment 1 and Segment 2, separated by a data downlink gap). Within this figure we marked five segments to which we applied the planetary-scale wave models (Segments 1abc and 2ab). The beginning and end times of the five segments were chosen somewhat arbitrarily, with the intent to identify overall representative segments of different lengths, to enable testing the planetary-scale wave model. Multiple other segments have been fitted, although not shown in the paper, with qualitatively identical results.

Figures~\ref{fig:Segment1a}, \ref{fig:Segment1b}, \ref{fig:Segment1c}, \ref{fig:Segment2a}, and \ref{fig:Segment2b} show our model fits to five lightcurve segments. In the upper panels of these figures, filled circles show the segment-normalized TESS photometric measurements (with colors reflecting the brightness of the object, i.e., the ordinate of the data points). The panel also includes the key parameters of the fits (periods and amplitudes of three sine waves). The lower panels show the residuals (data$-$model) in black filled circles, and the model itself (offset to 0.0 mean ordinate value).

 {We find that the lightcurve evolution over all five segments fitted is very well reproduced by the model. The standard deviation of the residuals remains very close to 0.01 for the fits, including the longest segment (100 hours, Segment 1c). It is instructive to see how this simple model succeeds in matching even segments with apparently complex lightcurve evolution: For example, in Segment 1b (Figure~\ref{fig:Segment1b}) in the course of forty hours the lightcurve evolves from a very low-amplitude, seemingly aperiodic section into a high-amplitude, flat-top section with deep, short minima. Or, for example, in Segment 1c (Figure~\ref{fig:Segment1c}, where in 100 hours the lightcurve goes from low-amplitude, sine-like variations to high-amplitude sine-like variations, before returning. In these cases, as in all other segments, the model reproduces the lightcurves well.}

 {A key question in the model fits is whether the fitted periods are, in fact, consistent with the Lomb-Scargle periodogram analysis' finding that the periods should form two groups (2.5~h and 5.28~h), specifically: How often do the fitted periods fall within the period ranges identified in the Lomb-Scargle periodogram, and how often are periods are inconsistent with those ranges? } 

 {We find, in fact, that all (fifteen) periods fitted by the model are in the ranges identified by the Lomb-Scargle periodogram analysis. The most deviant value is period 3 (P$_3$=5.94) in the shortest-fitted lightcurve, Segment 1a (Figure~\ref{fig:Segment1a}), but even this is relatively close to the group 4.75--5.75~h period group.}

 {In short, our lightcurve fits demonstrate that: A) Even long and complex sections of the lightcurve are very well reproduced by the plantetary scale wave model; and, B) The best-fit periods concentrate into the same two groups identified by the Lomb-Scargle analysis.}

%
\section{Discussion}
%

\subsection{Luhman 16A or Luhman 16B? Or both?}

{{An important challenge in interpreting the TESS data presented here is the fact that the TESS aperture includes both A and B components of the \luh{} system. Given the very small projected separation ($<0.3$\arcsec{}, \citealt{Bedin2017}) of the two components, it is not possible to identify the relative contributions of the two components to the TESS lightcurve based \textit{solely} on the TESS data.} 

{Nevertheless, a comparison of our results to published spatially resolved observations allows us to argue that the lightcurve presented here is (1) it is very likely dominated by a single component, and (2) that single component is \luh{} B. }}

{{First, our study of the Lomb-Scargle periodogram identifies a single very prominent periodicity (P=5.28~h) that is responsible for almost all of the short-term modulations. A much lower-amplitude  periodicity is also seen at P=6.9~h, which we tentatively attribute to the other component in the system. These data show that only a single component dominates the lightcurve: the alternative, i.e., both components contributing significantly, would require their rotational periods {\em and} amplitudes to match very closely (periods to within a few percent and amplitudes to a factor of $\sim$2). Such a coincidence in rotational period and amplitudes is extremely unlikely, leaving us with the conclusion that a single component (with P=5.28~h rotation period) dominates.}}

{{Second, the comparison to the literature measurements strongly suggests that this component is \luh{}~B. Although both components are variable \citep[e.g.,][]{Biller2013,Buenzli2015a}, the few spatially resolved studies all found B to display 2--3 higher amplitude in the near-infrared than A does. For example, spatially resolved HST Wide Field Camera 3 G102 and G141 spectrophotometry showed that, for wavelengths longer than 0.8~$\mu$m, component B has at least twice greater amplitude (4.6\%) than component A (2.2\%) \citep[][]{Buenzli2015a,Buenzli2015b}. Although continuous, spatially resolved, complete-rotation lightcurves are not available for wavelengths shorter than 0.8~$\mu$m, spatially resolved multi-epoch HST photometry of the system \citep[][]{Bedin2017} showed that, while components A and B have similar amplitude ($\sim$0.05~mag) in the F606W filter ($\lambda_c=606~nm$), in the redder F814W filter component B's amplitude ($\sim$0.12~mag) is more than twice of that of component A ($\sim$0.05~mag). 

In comparison, TESS's single band covers wavelengths between approximately 600-1,000~nm, with its central wavelength of $\lambda_c$=786\,nm. Therefore, although no complete, spatially resolved lightcurves are available for component A and B in TESS band, the data from multiple studies consistently and very strongly suggests that component B has at least 2 times higher band-integrated amplitude in the TESS-band than component A. In addition, the \luh{} system is a flux-reversal binary: the overall brighter primary \luh{}A is actually slightly fainter in the Y and J-bands than component B, further decreasing its relative contribution to the TESS-band.

Finally, there is also evidence from the periodicity of the objects: While A's period remain poorly constrained in the 5--8~h range \cite[e.g.,][]{Millar-Blanchaer2020}, B's amplitude has been established to be very close to 5.2~h:} \citet{Gillon2013} found a rotational period of 4.87$\pm$0.01h. Shortly after, \citet{Burgasser2014} fitted the lightcurve with a period of 5.05$\pm$0.1\,h, a longer period than estimated by \citet{Gillon2013}. \citet{Mancini2015} presented additional data supporting a longer rotational period (5.1$\pm$0.1\,h). {These rotational period measurements are completely consistent with the rotational period of 5.28~h we found in the Lomb-Scargle periodogram analysis (see Section~\ref{Section:PowerSpectrumAnalysis}), and very close to the value (P$\sim$5.4~h) found in the fits of the lightcurve segments via the planetary-scale wave model (Section~\ref{Section:LCAnalysis}).} }

{Therefore, we conclude that evidence strongly suggests that the lightcurve is dominated by modulations of a single component, and the facts that component has a P=5.28~h rotational period and that B has larger amplitude in the TESS-band strongly suggest that the component dominating the lightcurve is, in fact, component B. }

{We identify this period with the rotational period of 5.28~h derived from the Lomb-Scargle periodogram analysis (see Section~\ref{Section:PowerSpectrumAnalysis}), which we interpret as the rotational period of \luh{}B.} 
{Therefore, the P=5.28~h rotational period is fully consistent with recent literature values as well as the results of our periodogram analysis and the planetary-scale wave model fits.}

\subsection{The Rotational Periods and Inclinations of Luhman 16\,A and B}
\label{S:Inclinations}

{The combination of our finding of one dominant peak (P=5.28~h)  in the periodogram with literature-based identification of the B's rotational period of 5.2~h represents a step forward in understanding the \luh{}AB system. Based on our results B's rotational period is P=5.28~h (and it likely exhibits a small range of differential rotation). Furthermore, we tentatively attribute the 6.9~h peak in our periodogram to the rotational period of component A.}

{By combining the rotational periods for components A and B with measured radial rotational velocities for each component, we can also explore possible spin axis orientations in the system, i.e., constrain the inclinations of the two brown dwarfs. Assuming brown dwarf radii between 0.90 and 1.10~R$_{Jup}$ \citep[e.g.,][]{Burrows2001}, the equatorial rotational velocities expected from the rotational periods would be $v_{A,eq}=16.3-19.9$km/s and $v_{B,eq}=21.2-25.9$km/s. In contrast, \citet{Crossfield2014} measured projected rotational velocities of 17.6$\pm$0.1 km/s for \luh{}A and 26.1$\pm$0.2~km/s for \luh{}B. Thus, the observed $v\,sin(i)$ measurements are within 1$\sigma$ of the range of the equatorial rotational velocities calculated. This comparison is informative and shows that the measured rotational periods are fully consistent with the observed $v\,sin(i)$ values.} 

{Furthermore, our analysis suggests a close to equatorial viewing angle for both brown dwarfs. Based on our simple model, \luh{}A is viewed from an angles within 28$^\circ$ from its equatorial plane (i.e., $i>62^\circ$); and luh{}B is viewed almost exactly equatorially (within a few degrees, i.e., $i\simeq90^\circ$). Our analysis does not provide evidence for misaligned spin axes in the system; on the contrary, the two brown dwarfs may have very similarly aligned axes. As the \luh{} system's orbital motion has been closely monitored, it is possible to compare the spin axes of the brown dwarfs to that of their orbital spin axis: \citep{Bedin2017} found the orbital inclination to be 79.21$\pm$0.45$^\circ$, which suggests that the three rotational and orbital spin axes in the system may be well aligned.}

 {We note, that the power of the P=6.94~h peak, which we attribute to Luhman~16A, is significantly ($>10\times$) lower than that of the P=5.28~h peak, corresponding to Luhman 16~B's rotation. While \luh{}B's amplitude is lower, as discussed above, the relatively large difference between the two peaks in the TESS data is surprising. We speculate that this may suggest that the TESS observations occurred during a period when \luh{}A's amplitude was lower than usual. Future re-visits of the \luh{} system may help test this hypothesis and shed more light on the periodogram signal from \luh{}A. }  

{Finally, we note that a close-to-equatorial viewing angle is also consistent with the fact that the \luh{}B is among the highest-amplitude variable brown dwarfs known. High amplitudes are much more likely in systems seen in near-equatorial orientations \citep[e.g.,][]{Metchev2015,Vos2017}, as found here. }

\subsection{A Novel Look Into Brown Dwarf Atmospheres}

A stunning feature of the \luh{} lightcurve presented here is its complexity, i.e., non-periodic, evolving nature. Although complex lightcurve evolution has been reported for a number brown dwarfs \citep[e.g.,][]{Radigan2012,Apai2013, Gillon2013,Metchev2015,Karalidi2016}, most of these lightcurves have been too short and/or had too low signal-to-noise ratios to allow testing hypothetical models for lightcurve evolution. In contrast, the lightcurve shown here provides relative high signal-to-noise ratio and long temporal baseline. The only somewhat comparable lightcurve data is from the Spitzer Space Telescope Exploration Science program \textit{Extrasolar Storms} (see \citealt{Yang2016,Apai2017}). The \textit{Extrasolar Storms} data provided higher cadence and higher signal-to-noise ratio infrared (3.6 and 4.5~$\mu$m) photometry on six brown dwarfs, with overlapping \textit{HST} spectrophotometric snapshots. While the \textit{Extrasolar Storms} data covered up to $\sim$1,000 rotation periods for each target in eight visits, no visits covered more than four rotation periods. In contrast, the TESS data presented here provides a unique, quasi-continuous coverage of a single target over $\sim$100 rotation periods, thereby providing data on the previous unexplored 4--40 rotation timescale. Furthermore, the current dataset is the only high-precision, sustained monitoring carried out in the visible wavelength regime for a brown dwarf with continuously evolving lightcurves. Thus, the dataset provides a powerful comparison to the shorter-coverage infrared Spitzer monitoring data that are available for a larger set of objects.

\subsection{Lightcurve Evolution Model}

Based on the qualitative observation of the lightcurve (Figure~\ref{fig:Overview} and previous section) and the quantitative model fits described in Section~\ref{Section:LCAnalysis}, we conclude that the visual lightcurve of \luh{} is very similar in nature to the lightcurves observed in the infrared for this and other brown dwarfs \citep[][]{Buenzli2015a,Buenzli2015b,Karalidi2016,Apai2017}. Specifically, we identify four properties that are shared between the visual lightcurve of this object and the infrared lightcurves of other objects: 1) The lightcurves remain variable over long periods (years); 2) The lightcurve shape evolves, yet it displays  characteristic period, which is likely the rotational period of the object (as found in \citealt{Apai2017}); 3) In spite of the rapid evolution of the lightcurve, the amplitudes over rotational time-scales remain similar and characteristic to the object; 4) The lightcurves tend to be symmetric in the sense of similar amount of positive--negative features, in contrast to, for example, a situation in which a single positive feature appears periodically on an otherwise flat lightcurve, which would indicate a single bright spot in the atmosphere. 
Therefore, the qualitative similarity of the \luh{} LC to the data obtained for other L/T brown dwarfs in the \textit{Extrasolar Storms} program motivates the test of the planetary-scale waves model developed for those objects \citep{Apai2017}. In Figures~\ref{fig:Segment1a}--~\ref{fig:Segment2b} we applied the planetary-scale wave model to lightcurve segments of various lengths, ranging from 25~h (Segment~1a in Fig.~\ref{fig:Segment1a}), 40~h (Segment~1b, Fig.~\ref{fig:Segment1b}), 50~h (Segments 2a and 2b, Figs.~\ref{fig:Segment2a} and \ref{fig:Segment2b}), and 100~h (Segment 1c, Fig.~\ref{fig:Segment1c}). Each of these figures includes the key fit parameters (wave periods and amplitudes), as well as the data$-$model residuals.

 {We found that not only does the simple planetary-scale wave model provides an excellent fit to all lightcurve segments fitted, but that best-fit periods in the lightcurve segments correspond closely the periods identified in the periodogram analysis. Thus, the combination of these findings strongly supports the applicability of the planetary-scale waves to  Luhman~16B, i.e., strongly argues for the presence of multiple zones in the atmosphere with slightly different rotational periods -- in other words, it argues for zonal circulation.}

\subsection{Zonal Circulation in Luhman 16B}
\label{S:ZonalCirculation}

 {As revealed by the periodogram and by the lightcurve segment fits, a multitude of similar, but slightly different effective periods exists in the modulations of Luhman~16B. A very similar periodogram has been compiled from K2 (Kepler extended mission, \citealt[][]{Howell2014}) photometry of Neptune \citep[][]{Simon2016}, which was found to be similar to the periodogram derived from Spitzer lightcurves of two L/T transition brown dwarfs \citep[][]{Apai2017}. However, the data presented here provides a much higher quality periodogram in which much finer details are visible. The periodogram presented here reveals not just a simple split in the peak corresponding to the rotational period, but a multitude of peaks (Figure~\ref{fig:MultiplePeaks}). Furthermore, for the first time, it also provides similar insights into the weaker peak that corresponds to k=2 (half-period) waves (Figure~\ref{fig:MultiplePeaks2.5}). This "peak", too, is resolved to multiple similar peaks. }

 {The multi-peaked nature of the periodogram peaks demonstrates that Luhman~16B has zonal circulation, i.e., it sports latitudinal zones of eastward and westward jets. These high-speed jets rotating pro-grade and retrograde will modulate the rotational periods of any structures embedded in these zones: structures embedded in pro-grade (eastward jets) will display shorter rotational periods and structure embedded in westward jets will display longer rotational periods. Simulations \citep[e.g.,][]{Showman2013,Showman2019,Imamura2020} of giant planets and brown dwarfs, as well as observations of Jupiter \citep[e.g.,][]{Limaye1986,Porco2003} suggest that the equatorial jet is likely to be the fastest and also the widest jet, although its direction may oscillate between eastward and westward \citep[][]{Showman2019}.}

 {With these insights in mind, we interpret Figure~\ref{fig:MultiplePeaks} in the following way: The strongest period detected (P=5.29~h) likely corresponds to the period of the equatorial jet, i.e., it is slightly different from the true rotational period (interior) of \luh{}B. Peaks that are shorter than the 5.25~h are likely emerging from structures embedded in prograde (eastward) jets, while structures embedded in retrograde (westward) jets are responsible for the peaks at periods longer than $\sim$5.29~h. Whether the equatorial jet itself is prograde or retrograde cannot, at this point, be determined.}

 {We note that the number of peaks fitted to the periodogram do not correspond directly to the number of jets presents in the atmosphere, as any number of jets with similar speeds will appear as a single peak. Nevertheless, the number of peaks may place a lower limit on the number of zones, under three reasonable assumptions: (1) wind speeds in a given zone are constants, (2) wind speeds are approximately symmetric to the equator, (3) there are no transition regions between the zones. Such a simple model would suggest that Luhman~16B harbors at least six distinct zones in velocity space -- or, given equatorial symmetry -- a total twelve zones (or six band/belt pairs).}

 {Although the model described above is useful as a first approximation, we note that the latitudinal wind velocity distribution in real objects is unlikely to be this simple. In fact, Cassini observations of wind speeds in Jupiter \citep[][]{Limaye1986,Porco2003}, as well as circulation models of brown dwarfs show no sudden breaks in the latitudinal wind speed distribution, but rather smooth transitions. 
Thus, models based on a number of belts with distinct velocities (or effective periods) have only limited validity. We further explore these considerations with a simple model in Section~\ref{S:JupiterComparison} and note, that future studies may attempt more direct comparisons between circulation models and periodograms.}

 {Nevertheless, our data does allow the exploration of the relative wind velocities presented in Luhman~16B. Although the exact period range in which peaks are present is not well determined, we find evidence for at least a 22\% relative period range (Figure~\ref{fig:MultiplePeaks}). Given the rotational period and the typical diameter of a brown dwarf, the maximum wind speed \textit{difference} corresponds to about 4.5~km/s. Interestingly, this value is about an order of magnitude greater than that measured in Jupiter or predicted by baseline brown dwarf models \citep[][]{Showman2019}. 
 We note that, perhaps, this discrepancy may be resolved if the structures seen are not just passive tracers embedded in the jet systems, but truly perturbations that propagate within the jet systems. For example, radiative--convective feedbacks have been proposed to introduce periodic cloud formation/dissolution \citep[][]{Tan2017}. If such perturbations do exist, then the measured velocities would correspond to the sum of the jet velocities and the propagation speed of the perturbations, thus significantly narrowing the gap between predicted wind speeds and observed periods.}
 
 \added{It is also useful to place the observed wind speed \textit{range} of 4.5~km/s in the context of the recent study by \citet[][]{Allers2020}, who derived the wind speed in a brown dwarf relative to its interior by comparing its near-infrared rotational period to its radio-based rotational period. This study found an eastward wind speed of 650$\pm$310~m/s for a rapidly-rotating (P=1.77$\pm$0.04~h) T6 spectral type brown dwarf. Considering the uncertainties, this is about a factor of three lower than the wind speed \textit{range} we derived for the slower-rotating and hotter \luh{}B. This difference, however, should not be surprising as the radio--near-infrared period comparison is likely measuring the relative speed between the \textit{dominant jet} and the brown dwarf's interior, while our measurements capture the velocity differences between the fastest eastward and the fastest westward jets, and thus are expected to be higher.} 

\subsection{Comparison to Jupiter Model}
\label{S:JupiterComparison}

To place the \luh{}B periodogram results in context, we will now explore how Jupiter -- if observed similarly to \luh{}B -- may compare to them. Jupiter is, of course, not a perfect analog to \luh{}B, as they differ in effective temperature, surface gravity, and -- very likely -- in composition. Yet, their general atmospheric circulation may share similarities and a comparison may provide some insights. There is no identical dataset available, and we attempt here to build an admittedly simple model for Jupiter, aiming at a qualitative, exploratory comparison. 

Using wind speed measurements (as a function of latitude) for Jupiter, we will calculate a histogram of the effective period across the atmosphere of the planet. The fundamental, simplifying assumption we make here is that every visible atmospheric element contributes to the signal that emerges from Jupiter equally, adding its own periodicity. In other words, the larger fraction of the atmosphere has a given wind speed, the stronger the effective period that correspond to that wind speed value will appear in the histogram. 

\begin{figure}[h]
\begin{center}
\includegraphics[width=1.0\linewidth]{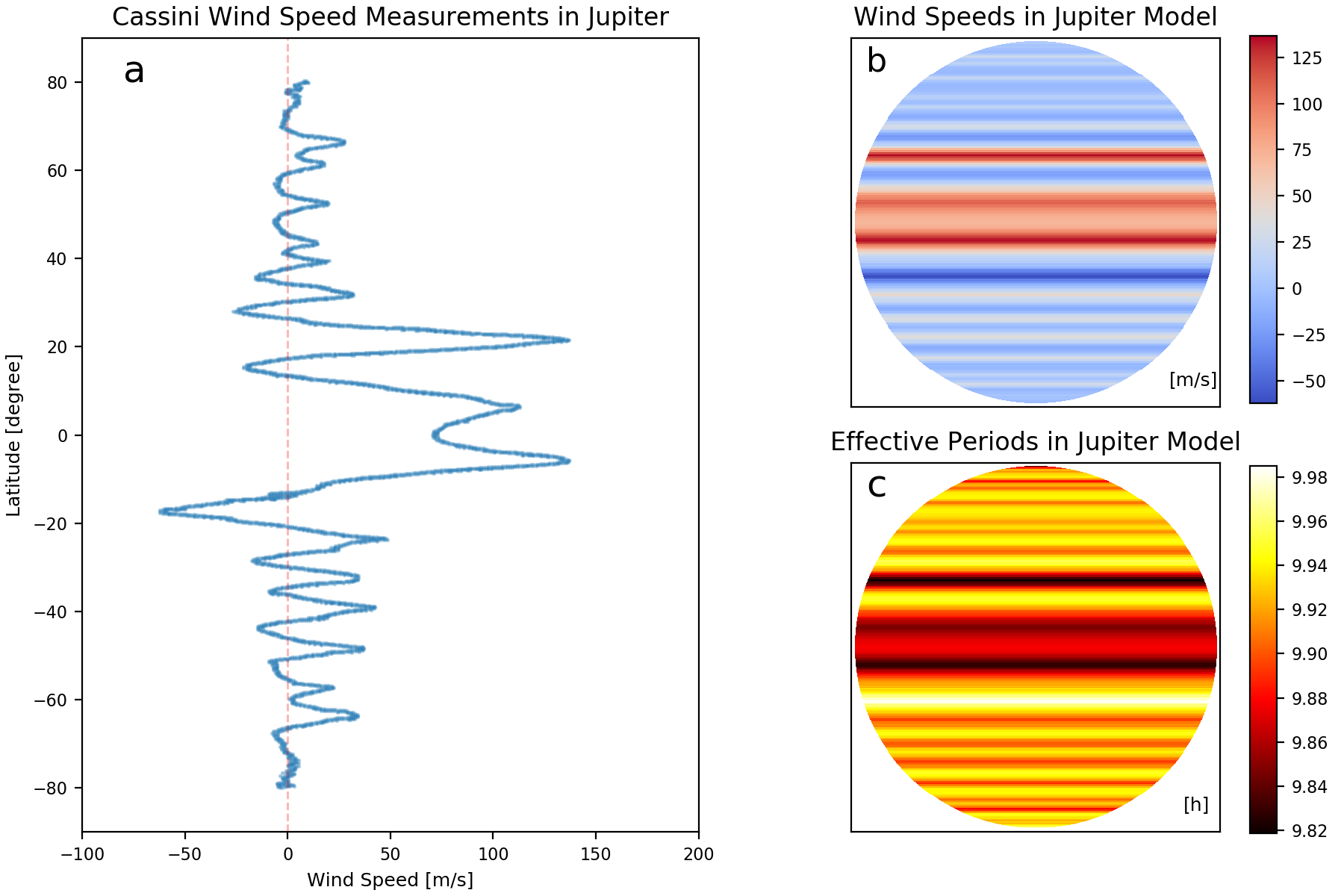}
\caption{A simple model for Jupiter to support comparison to the \luh{}B periodogram. \textit{(a)} Wind speeds measured as a function of latitude in Jupiter by the Cassini team \citep[][]{Porco2003}. \textit{(b)} The observed wind speeds mapped to a spherical disk. \textit{(c)} Effective periods calculated from the sum of the Jupiter's bulk rotation and the latitudinally varying wind speeds. \label{fig:JupiterModel}}
\end{center}
\end{figure}

Panel \textit{a} of Figure~\ref{fig:JupiterModel} shows the wind speed values as a function of latitude, as measured by the Cassini mission \citep[from][]{Porco2003}. The highest wind speeds ($\sim$130 m/s and $\sim-70$~m/s) were observed at low latitudes, close to the equator, while more moderate wind speeds (typically between $-10$~m/s and $+50$~m/s) were observed at mid- to high latitudes. We note that a very similar wind speed pattern was also measured in Voyager 1 and Voyager 2 image pair differences \citep[][]{Limaye1986} about two decades earlier, therefore the wind speed pattern shown is characteristic to Jupiter's current atmospheric circulation state. As our goal is to translate the wind speed pattern into a histogram of effective periods, we map the wind speeds ($v_{\mathrm{wind}}$) observed onto a simulated disk of Jupiter (approximating its shape with a sphere), as shown in Panel \textit{b} of Figure~\ref{fig:JupiterModel}. We then calculate an effective period map for our synthetic Jupiter, by adopting an equatorial rotational period of $P_\mathrm{eq}=$9$^\mathrm{h}$56$^\mathrm{m}$, a radius of 1 Jupiter radius, and calculating, for each atmospheric element, the rotational period that emerges from the combination of the local wind speed and the bulk rotation of the planet:
$$P_{\mathrm{eff}}(\lambda)= {2 \pi R_{\mathrm{Jup}} * \cos(\lambda) \over v_{\mathrm{rot}} + v_{\mathrm{wind}}(\lambda) } \mathrm{,}$$

where $\lambda$ denotes the latitude, $R_{\mathrm{Jup}}$ is 1 Jupiter radius, $v_{\mathrm{rot}}=2 \pi R_{\mathrm{Jup}} / P_{\mathrm{eq}}$ is the rotational velocity corresponding to the bulk rotation of Jupiter.  
 The resulting map is shown in Panel \textit{c}. In this synthetic map, rotational periods range from 9$^\mathrm{h}$49$^\mathrm{m}$ to 9$^\mathrm{h}$59$^\mathrm{m}$. 

\begin{figure}[h]
\begin{center}
\includegraphics[width=1.0\linewidth]{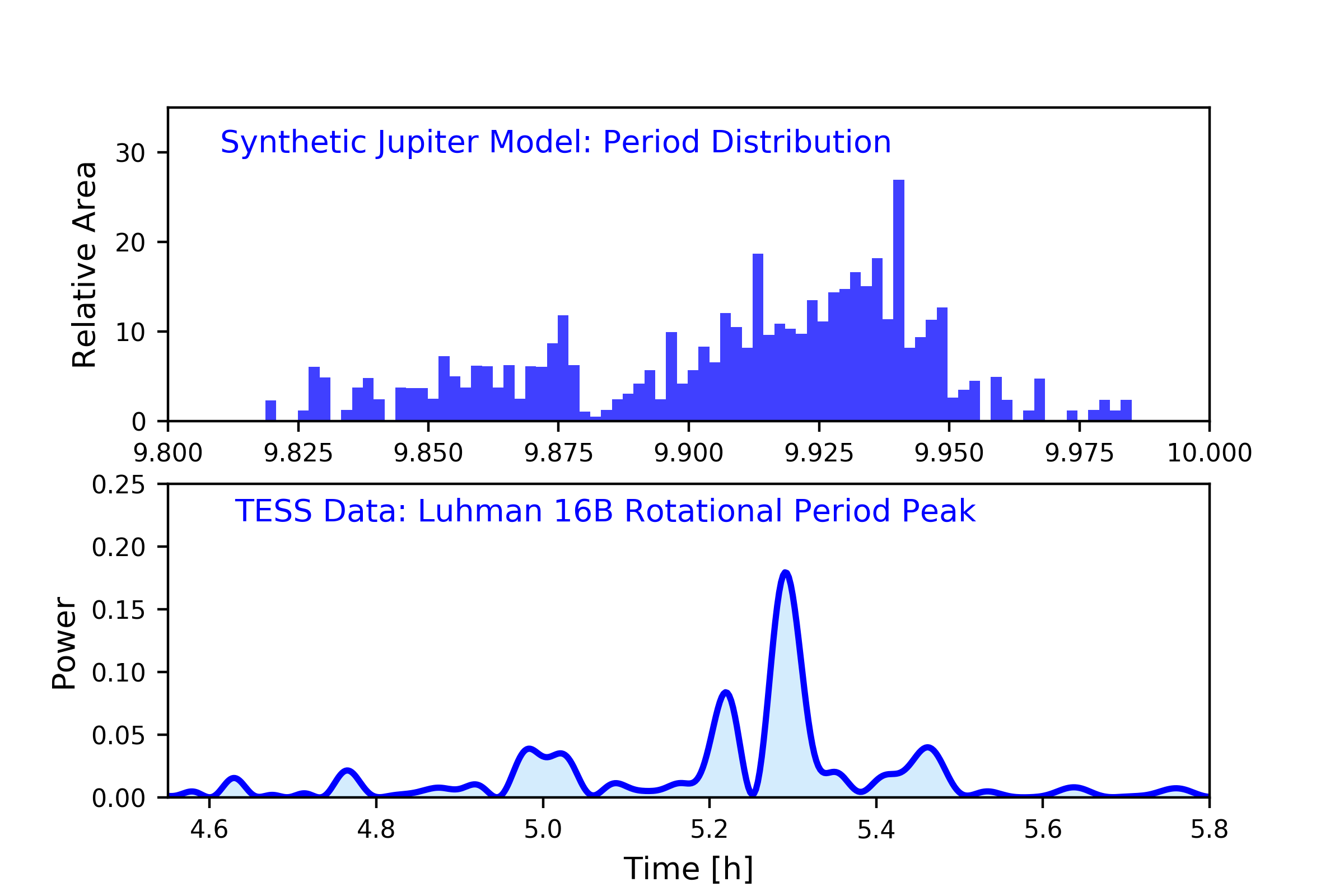}
\caption{\textit{Top panel:} Histogram of the effective periods in our Jupiter model (Figure~\ref{fig:JupiterModel}, panel c). \textit{Bottom panel:} The multi-peaked structure in the periodogram of the \luh{}B rotational peak (P=5.28~h). In spite of the important differences between Jupiter and \luh{}B and the simplicity of our Jupiter model, there is an overall qualitative similarity between the distributions of effective periods, supporting the the conclusion that \luh{}B has zonal circulation, as Jupiter does. \label{fig:JupiterComparison}}
\end{center}
\end{figure}

We will now contrast the predictions of this very simple Jupiter model to the observations of \luh{}B. In the top panel of Figure~\ref{fig:JupiterComparison} we plot the histogram of the effective periods of the photospheric elements in our Jupiter model. In the lower panel of the same figure we plot, for comparison, the Lomb-Scargle periodogram results for the primary rotational peak (P=5.28~h) of \luh{}B (from Section~\ref{s:Peaks}). We find that, in spite of the simplicity of our Jupiter model and the limited amount of information on \luh{}B, there are some interesting qualitative similarities between the results. 
First, in both cases there is a range of effective periods present in the vicinity of the objects' rotational periods. Second, the distributions are asymmetric: there are longer tails toward shorter periods than toward longer periods, suggesting that larger parts of the atmosphere are characterized by high-speed \textit{prograde} jets than by retrograde ones, and that the prograde jets have higher velocity. Third, in both distribution there appears to be evidence for distinct peaks, i.e., there are likely some typical, characteristic jet speeds rather than a smooth distribution of wind speeds without peaks. 
The comparison also reveals two differences: First, the relative range of periods predicted for Jupiter is (as mentioned in Section~\ref{S:ZonalCirculation}) is about an order-of-magnitude lower than that seen in \luh{}B. Second, the periodogram of \luh{}B shows more distinct period peaks than those seen in our synthetic Jupiter model. The latter point may simply be a result of our simple model, where we assumed that each atmospheric element contributes equally to the periodogram. In reality, Jupiter's thermal emission is very unevenly distributed and opacity holes -- infrared bright spots -- are strongly correlated with the atmospheric circulation \citep[e.g.,][]{Showman2000}. Due to this correlation between wind speed and local thermal infrared brightness, it is very likely that less simplistic models of Jupiter's periodogram will display more prominent peaks.

The former difference, however, between the relative range of periods in Jupiter and the \luh{}B lightcurve is very unlikely to be the result of our simplistic model, as this aspect of the prediction is the direct consequence of the measured wind speeds. The difference between \luh{}B and Jupiter, in terms of their relative period range, does suggest that an additional, different process is occurring in the brown dwarf than in Jupiter. The propagation of planetary-scale waves \citep[][]{Apai2017}, and/or waves caused by radiative-convection feedbacks and cloud formation \citep[][]{Tan2017} may be candidates to explain the different relative period ranges.

\added{Our model assumes a perfectly equatorial viewing angle, which is -- as explained in Section~\ref{S:Inclinations} -- likely very close to \luh{}B's orientation. Nevertheless, we will briefly address here how the predicted period distribution may change for more inclined (10--30$^\circ$) viewing angles. Unlike for a Doppler-shift-based measurement, the change in the viewing angle for rotational period-based measurements will not result directly in the compression of the deduced velocity/period differences: This is because the measurements probe the rotational periods of different latitudes, which are independent of the viewing angle. The viewing angle will, however, lead to a second-order effect. Specifically, it will impact which latitude range is visible in the atmosphere. This, in turn, will weaken the power of  the periodogram peaks that correspond to the less-visible latitudes, while leaving the power of other peaks essentially unchanged. Therefore, we conclude that our model's fundamental predictions are mostly insensitive to the exact viewing angle (as long as they are generally close to equatorial, as expected for \luh{}B), and that slight inclinations would impact the periodogram power of the peaks but not their periods or the range of periods observed.}

\begin{figure}[h]
\begin{center}
\includegraphics[width=1.0\linewidth]{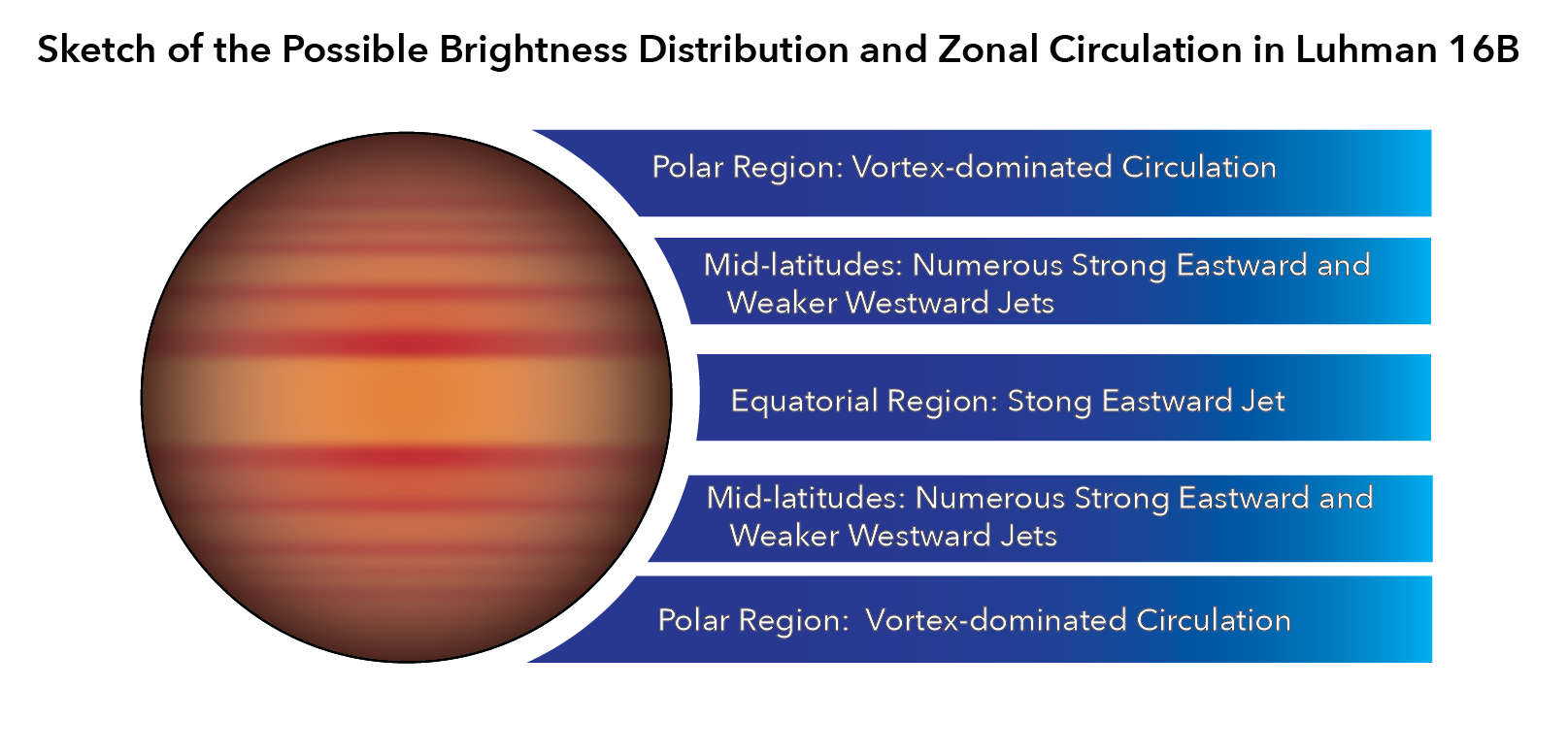}
\caption{Sketch of the possible appearance of \luh{}B, based on the emerging evidence. Zonal circulation models and comparison to Jupiter suggests that low-latitude regions are dominated by the fastest jets, and that wind speeds at mid-latitude are significantly lower. Circulation at the polar regions is likely to be vortex- and not jet-dominated. Cloud cover is likely to be correlated with the atmospheric circulation. \label{fig:Sketch}}
\end{center}
\end{figure}

Thus, given the above, we conclude that, in spite of an important quantitative difference between the simulated Jupiter periodogram and the observed \luh{}B periodogram, there are multiple qualitative similarities that support the interpretation that Jupiter and \luh{}B share similarities between their atmospheric circulation. Therefore, we take the qualitative similarities as further support for the conclusion that \luh{}B displays zonal circulation. Figure~\ref{fig:Sketch} shows a qualitative sketch that illustrates the possible appearance of \luh{}B.

\subsection{Long-time scale evolution}
\label{S:LongPeriodEvolution}

 {Given our analysis, we have now viable models and explanations for the natures of the short-period peaks (P=2.5~h, P=5.28~h, P=6.94~h) in the \luh{}AB system. However, the TESS lightcurve also includes a long-period component, which appears in the Lomb-Scargle periodogram as a P=90.8~h peak, although its exact period is not well determined. Our analysis of spacecraft positional jitter (Section~\ref{S:TESSPhotometry}) and background star variability (Section~\ref{S:BackgroundStars}) strongly argue against such contamination and, therefore, for the genuine nature of the observed intensity variations. We will now briefly speculate about the possible nature of these variations.}

 {The overall amplitude of the observed long-term modulations is about 10\%, almost twice as large as the amplitude of the short-term modulations (mainly P=5.28~h). Given that we identified the rotational periods of both components A and B, and that these are 6.94~h and 5.28~h, respectively, we can establish that the long-period variability -- regardless of its precise period -- is more than an order of magnitude longer than the rotational periods. This observation means that the part of the atmosphere that is introducing the changes is \textit{visible throughout a rotational cycle}. }

 {Given these observations, we speculate that the long-period changes observed here emerge from one (or both) of the following sources: (a) global atmospheric evolution affecting both hemispheres in a coordinated manner; (b) permanently visible polar regions. }

 {With the information at hand we cannot distinguish between these two scenarios nor determine whether the long-period modulations emerge from component A or B (or both). Nevertheless, we briefly discuss both scenarios and the possible source. Scenario \textsc{I} -- global evolution of atmospheric properties -- is interesting as such a global evolution (e.g., cloud cover, atmospheric chemistry, or oscillations in pressure-temperature profile) would seemingly naturally explain the long-period modulations. However, the $\sim$90~h timescale is likely too short to be explained by such a global evolution. Disturbances that propagate with the sound speed ($\sim$1km/s) may circle a brown dwarf in $\sim$100~h. However, global changes are likely to propagate at significantly sub-sonic velocities, especially in the latitudinal direction, where -- presumably -- circulation zones must be crossed. An example for long-period oscillations in brown dwarfs, as predicted by \citet[][]{Showman2019}, are quasi-biennial oscillations (QBOs). Simulations predict that such oscillations would have a timescale of 1,000 to 10,000 Earth days, i.e., about three orders of magnitude longer than the P$\sim$90~h period observed in our lightcurve. Therefore, we consider global atmospheric evolution to be an unlikely cause of the P$\sim$90~h lightcurve evolution.}

 {In scenario \textsc{II} the permanently visible polar regions of the brown dwarf evolve. We note that no particular inclination is required for the polar regions to be permanently visible (as even an equatorial view will probe the polar regions), but brown dwarfs viewed more pole-on will also allow circumpolar regions to be observed. There are reasons to think that the polar regions will be morphologically distinct from the low- and mid-latitude regions: such difference exists for both Saturn and Jupiter. Furthermore, it is predicted by three-dimensional circulation models for brown dwarfs, too: While the low- to mid-latitude regions are typically dominated by jets, the polar regions are dominated by wave dynamics and characterized, in appearance by vortices. The  latitude of transition between the jet- and the vortex-dominated circulation depends primarily on the dynamical friction, wind speeds, and rotational speed. The polar regions are smaller in size than the two hemisphere area scenario \textsc{I} required to alter; and the polar regions are not delineated by high-speed atmospheric jet boundaries. Therefore, naively, these regions appear to be more likely candidates for the source of the long-period evolution. If so, then the changes observed may correspond to rearrangements of vortices, i.e., possibly the evolution of large polar storms. We further speculate that, if indeed we are observing polar storm systems, these may more likely be observed on Luhman~16A, which appears to be observed at a slightly inclined viewing angle (Section~\ref{S:Inclinations}).}

 {Although our study cannot determine the nature and source of the long-term evolution of the lightcurve, we note that the hypothesis that the long-term evolution probes permanently or quasi-permanently visible polar regions could be testable with future monitoring of Luhman~16AB. In addition, a likely prediction of this hypothesis is an anti-correlation between short-term and long-term variability: Sources seen closer to equatorial viewing angles may display greater rotational modulations, while sources viewed closer to the polar angles would be dominated by the slower-evolving polar storm systems. If this hypothesis is confirmed, it will become possible to elevate our understanding of atmospheric circulation of brown dwarfs by studying both their jet-dominated zonal circulation and their vortex-dominated polar regions.}

\subsection{State of the Art and Outlook}

The TESS LC presented here -- and potential future, even higher-quality data -- will allow the   {detailed tests of brown dwarf circulation models}. Most importantly, the quasi-continuous nature of the TESS LC allows the study of the time-evolution of the wave components present in a way no other existing dataset does. The exploratory Spitzer monitoring by \citet{Apai2017} provided LC segments that were too far apart to sample the LC evolution between segments, revealing that the timescale of wave component evolution is shorter than the 20--100 rotations that separated the Spitzer LC segments. 

 {The data presented here strongly supports the general model put forward by \citet[][]{Apai2017}}. The current TESS dataset  provides a relatively well-sampled, quasi-continuous  {lightcurve that enabled the identification of different circulation zones within Luhman~16B. This dataset sets the stage for direct tests of large-scale circulation models \citep[e.g.,][]{ZhangShowman2014,Tan2017,Tan2019} through the comparisons of not the lightcurves, but the periodogram generated from the simulated data and from the observations.}

 {In Section~\ref{S:LongPeriodEvolution} we also proposed the hypothesis that the long period evolution of the lightcurve probes the evolution of the vortex-dominated polar regions. This is a testable hypothesis. If confirmed, the combination of the short-term and long-term variability would open a window to both the jet-dominated and the vortex-dominated regions of brown dwarf atmospheres.}

In fact, future TESS observations have {the potential to further refine the periods present in the lightcurve and also to constrain the long-term ($\sim$1,000 rotations) evolution of the periodogram.} 
Given that \luh{} is a uniquely bright variable brown dwarf, it is likely that it will remain the best target for future TESS observations. We note, furthermore, that to maximize the science return of the unique TESS data, simultaneous multi-wavelength lightcurves are required. Such multi-wavelength datasets have been used to constrain, for example, the likely latitudinal width of the zonal circulation-driven banded structure \citep[][]{Apai2017} or, in \citet[][]{Buenzli2012}, simultaneous HST and Spitzer observations were used to uncover large-scale latitudinal-vertical organization in a late T-type brown dwarf. A possible re-visit of \luh{} by TESS will be a unique opportunity to obtain simultaneous multi-wavelength observations (even if much shorter in coverage).

\added{Finally, we note that detailed and high-precision observations of the effective periods of the jet-dominated regions, such as those presented here, may be complemented by radio-wavelength period measurements that probe the rotation rate of the interior of the brown dwarfs. In a recent study \citep[][]{Allers2020} demonstrated the potential of comparing radio periods to near-infrared rotation periods and, from their difference, deduced an eastward wind speed of 650$\pm$310~m/s. The TESS observations presented here do not probe the rotation rate of the interior and, thus, lack an absolute reference frame for wind speed measurements. Complementing them in the future with radio observations will enhance their diagnostic power for studies of atmospheric dynamics in Luhman~16B. This technique, of course, can also be extended for other, sufficiently bright brown dwarfs.}

\section{Conclusions}

Our study presents TESS lightcurves of \luh{} that cover about 22 days. The key findings of our study are as follows:

1) The observations cover approximately 100 rotations of \luh{}, during which the lightcurve remained non-constant, i.e., exhibiting significant rotational modulations.

2) The lightcurve of \luh{} displays clear evolution during the observations: The lightcurve evolves rapidly and continuously, even over timescales of 1--2 rotational periods. 

3) We find peak-to-peak variability amplitudes over timescale of a single assumed rotational period to be $\sim$4\% or less, but detect brightness differences as large as $\sim$10\% over longer baselines (100~h).

4) Based on Lomb-Scargle periodogram analysis we identify the rotational period of the component that dominates the short-term modulations as 5.28~h. {We also identify peaks at 2.5~h and 6.94~h, as well as a long-period peak with a period around 91~h.}

5) Based on the periodogram analysis, we conclude that the lightcurve is very likely dominated by one of the binary components, but not both.

6)  {We find strong evidence that the 2.5~h and 5.28~h peaks emerge from Luhman~16B, while the 6.94~h likely corresponds to the rotation period of Luhman~16A. The 2.5~h peaks likely correspond to k=2 (half-period) waves, as seen in other brown dwarf lightcurves \citep[]{Apai2017}.}

 {7) Our analysis demonstrates that, in the periodogram, both the 2.5~h and the 5.28~h peaks consist of multiple adjacent peaks, a robust signature of the combined effects of differential rotation and planetary-scale waves. The fitted periods demonstrate an at least 20\% rotational period range in Luhman~16~B, likely emerging from the combination of strong differential rotation and high-speed winds.}

8) We find that the derived rotational periods are fully consistent with observed $v\,\sin(i)$ values and that their combination suggests closely equatorial viewing angles (within $25^\circ$ for A and within a few degrees for B).

9) The lightcurve evolution observed in \luh{} is   {very well fit by} the planetary-scale model proposed by \cite{Apai2017}. 

10)  {We present a simple model for Jupiter, based on Cassini wind speed measurements, and show that the predicted periodogram is qualitatively similar to that of \luh{}B. This finding further supports the conclusion that \luh{}B's atmosphere is shaped by zonal circulation.}

9) {We cannot determine whether the long-period ($\sim$90.8~h) variability originates in Luhman 16A or B, but propose that these modulations emerge from the (quasi-)permanently visible vortex-dominated polar regions. This is a testable hypothesis that may pave the way toward a more complete understanding of atmospheric circulation in brown dwarfs.}

\added{A TESS re-visit of the \luh{} system -- currently scheduled for March-April 2021 -- will provide a valuable dataset to contrast to that presented here, allowing the comparison of the nature of the lightcurve evolution over a baseline of two years. } 
Given our study and the recent results by \citet[][]{Millar-Blanchaer2020}, we can now conclude that \textit{both} brown dwarfs in the \luh{} system display  zonal circulation. Our study opens up the possibility, for the first time, to directly compare predictions of circulation models to observations via periodogram analysis of long-term monitoring datasets of brown dwarfs.

\section{Acknowledgments}
We are grateful to the referee whose suggestions have fundamentally changed and enhanced this study. LRB acknowledges support by MIUR under PRIN program \#2017Z2HSMF. DN acknowledges support from the French Centre National d'Etudes Spatiales (CNES). This paper includes data collected by the TESS mission. Funding for the TESS mission is provided by the NASA Explorer Program. This study made use of data obtained in the Hubble Space Telescope programs GO-13748 and GO-14330. 
This work makes also use of results from the European Space Agency (ESA) space mission Gaia. 
Gaia data are being processed by the Gaia Data Processing and Analysis Consortium (DPAC). 
Funding for the DPAC is provided by national institutions, in particular the institutions participating in the Gaia MultiLateral Agreement (MLA). 
The Gaia mission website is \texttt{https://\-www.cosmos.esa.int/\-gaia}. 
The Gaia archive website is \texttt{https://\-archives.esac.esa.int/\-gaia}. 

\vspace{5mm}
\facilities{TESS, Cassini, Hubble Space Telescope}

\software{astropy \citep{astropy2013}. IDL-implementation of Genetic Algorithm (\texttt{https://\-www.ncnr.nist.gov/\-staff/\-dimeo/\-idl\_programs.html}). 
Planetary-scale wave model \citep{Apai2017}. SWarp \citep{Bertin2002}. VARTOOLS \citep{HartmanBakos2016}. ASTROMETRY.NET \citep{lang2010}. AstroPy \citep{Czesla2019}.}

\bibliography{apairefs}{}

\begin{thebibliography}{}
\expandafter\ifx\csname natexlab\endcsname\relax\def\natexlab#1{#1}\fi
\providecommand{\url}[1]{\href{#1}{#1}}
\providecommand{\dodoi}[1]{doi:~\href{http://doi.org/#1}{\nolinkurl{#1}}}
\providecommand{\doeprint}[1]{\href{http://ascl.net/#1}{\nolinkurl{http://ascl.net/#1}}}
\providecommand{\doarXiv}[1]{\href{https://arxiv.org/abs/#1}{\nolinkurl{https://arxiv.org/abs/#1}}}

\bibitem[{{Allers} {et~al.}(2020){Allers}, {Vos}, {Biller}, \&
  {Williams}}]{Allers2020}
{Allers}, K.~N., {Vos}, J.~M., {Biller}, B.~A., \& {Williams}, P. K.~G. 2020,
  Science, 368, 169, \dodoi{10.1126/science.aaz2856}

\bibitem[{{Ammons} \& {Garcia}(2019)}]{Ammons2019}
{Ammons}, S.~M., \& {Garcia}, V. 2019, in American Astronomical Society Meeting
  Abstracts, Vol. 233, American Astronomical Society Meeting Abstracts \#233,
  114.07

\bibitem[{{Apai} {et~al.}(2013){Apai}, {Radigan}, {Buenzli}, {Burrows}, {Reid},
  \& {Jayawardhana}}]{Apai2013}
{Apai}, D., {Radigan}, J., {Buenzli}, E., {et~al.} 2013, \apj, 768, 121,
  \dodoi{10.1088/0004-637X/768/2/121}

\bibitem[{{Apai} {et~al.}(2017){Apai}, {Karalidi}, {Marley}, {Yang}, {Flateau},
  {Metchev}, {Cowan}, {Buenzli}, {Burgasser}, {Radigan}, {Artigau}, \&
  {Lowrance}}]{Apai2017}
{Apai}, D., {Karalidi}, T., {Marley}, M.~S., {et~al.} 2017, Science, 357, 683,
  \dodoi{10.1126/science.aam9848}

\bibitem[{{Astropy Collaboration} {et~al.}(2013){Astropy Collaboration},
  {Robitaille}, {Tollerud}, {Greenfield}, {Droettboom}, {Bray}, {Aldcroft},
  {Davis}, {Ginsburg}, {Price-Whelan}, {Kerzendorf}, {Conley}, {Crighton},
  {Barbary}, {Muna}, {Ferguson}, {Grollier}, {Parikh}, {Nair}, {Unther},
  {Deil}, {Woillez}, {Conseil}, {Kramer}, {Turner}, {Singer}, {Fox}, {Weaver},
  {Zabalza}, {Edwards}, {Azalee Bostroem}, {Burke}, {Casey}, {Crawford},
  {Dencheva}, {Ely}, {Jenness}, {Labrie}, {Lim}, {Pierfederici}, {Pontzen},
  {Ptak}, {Refsdal}, {Servillat}, \& {Streicher}}]{astropy2013}
{Astropy Collaboration}, {Robitaille}, T.~P., {Tollerud}, E.~J., {et~al.} 2013,
  \aap, 558, A33, \dodoi{10.1051/0004-6361/201322068}

\bibitem[{{Bedin} {et~al.}(2017){Bedin}, {Pourbaix}, {Apai}, {Burgasser},
  {Buenzli}, {Boffin}, \& {Libralato}}]{Bedin2017}
{Bedin}, L.~R., {Pourbaix}, D., {Apai}, D., {et~al.} 2017, \mnras, 470, 1140,
  \dodoi{10.1093/mnras/stx1177}

\bibitem[{{Benatti} {et~al.}(2019){Benatti}, {Nardiello}, {Malavolta},
  {Desidera}, {Borsato}, {Nascimbeni}, {Damasso}, {D'Orazi}, {Mesa}, {Messina},
  {Esposito}, {Bignamini}, {Claudi}, {Covino}, {Lovis}, \&
  {Sabotta}}]{Benatti2019}
{Benatti}, S., {Nardiello}, D., {Malavolta}, L., {et~al.} 2019, \aap, 630, A81,
  \dodoi{10.1051/0004-6361/201935598}

\bibitem[{{Bertin} {et~al.}(2002){Bertin}, {Mellier}, {Radovich}, {Missonnier},
  {Didelon}, \& {Morin}}]{Bertin2002}
{Bertin}, E., {Mellier}, Y., {Radovich}, M., {et~al.} 2002, Astronomical
  Society of the Pacific Conference Series, Vol. 281, {The TERAPIX Pipeline},
  ed. D.~A. {Bohlender}, D.~{Durand}, \& T.~H. {Handley}, 228

\bibitem[{{Biller} {et~al.}(2013){Biller}, {Crossfield}, {Mancini}, {Ciceri},
  {Southworth}, {Kopytova}, {Bonnefoy}, {Deacon}, {Schlieder}, {Buenzli},
  {Brandner}, {Allard}, {Homeier}, {Freytag}, {Bailer-Jones}, {Greiner},
  {Henning}, \& {Goldman}}]{Biller2013}
{Biller}, B.~A., {Crossfield}, I. J.~M., {Mancini}, L., {et~al.} 2013, \apjl,
  778, L10, \dodoi{10.1088/2041-8205/778/1/L10}

\bibitem[{{Buenzli} {et~al.}(2015{\natexlab{a}}){Buenzli}, {Marley}, {Apai},
  {Saumon}, {Biller}, {Crossfield}, \& {Radigan}}]{Buenzli2015b}
{Buenzli}, E., {Marley}, M.~S., {Apai}, D., {et~al.} 2015{\natexlab{a}}, \apj,
  812, 163, \dodoi{10.1088/0004-637X/812/2/163}

\bibitem[{{Buenzli} {et~al.}(2015{\natexlab{b}}){Buenzli}, {Saumon}, {Marley},
  {Apai}, {Radigan}, {Bedin}, {Reid}, \& {Morley}}]{Buenzli2015a}
{Buenzli}, E., {Saumon}, D., {Marley}, M.~S., {et~al.} 2015{\natexlab{b}},
  \apj, 798, 127, \dodoi{10.1088/0004-637X/798/2/127}

\bibitem[{{Buenzli} {et~al.}(2012){Buenzli}, {Apai}, {Morley}, {Flateau},
  {Showman}, {Burrows}, {Marley}, {Lewis}, \& {Reid}}]{Buenzli2012}
{Buenzli}, E., {Apai}, D., {Morley}, C.~V., {et~al.} 2012, \apjl, 760, L31,
  \dodoi{10.1088/2041-8205/760/2/L31}

\bibitem[{{Burgasser} {et~al.}(2013){Burgasser}, {Sheppard}, \&
  {Luhman}}]{Burgasser2013}
{Burgasser}, A.~J., {Sheppard}, S.~S., \& {Luhman}, K.~L. 2013, \apj, 772, 129,
  \dodoi{10.1088/0004-637X/772/2/129}

\bibitem[{{Burgasser} {et~al.}(2014){Burgasser}, {Gillon}, {Faherty},
  {Radigan}, {Triaud}, {Plavchan}, {Street}, {Jehin}, {Delrez}, \&
  {Opitom}}]{Burgasser2014}
{Burgasser}, A.~J., {Gillon}, M., {Faherty}, J.~K., {et~al.} 2014, \apj, 785,
  48, \dodoi{10.1088/0004-637X/785/1/48}

\bibitem[{{Burrows} {et~al.}(2001){Burrows}, {Hubbard}, {Lunine}, \&
  {Liebert}}]{Burrows2001}
{Burrows}, A., {Hubbard}, W.~B., {Lunine}, J.~I., \& {Liebert}, J. 2001,
  Reviews of Modern Physics, 73, 719, \dodoi{10.1103/RevModPhys.73.719}

\bibitem[{{Crossfield} {et~al.}(2014){Crossfield}, {Biller}, {Schlieder},
  {Deacon}, {Bonnefoy}, {Homeier}, {Allard}, {Buenzli}, {Henning}, {Brandner},
  {Goldman}, \& {Kopytova}}]{Crossfield2014}
{Crossfield}, I.~J.~M., {Biller}, B., {Schlieder}, J.~E., {et~al.} 2014, \nat,
  505, 654, \dodoi{10.1038/nature12955}

\bibitem[{{Czesla} {et~al.}(2019){Czesla}, {Schr{\"o}ter}, {Schneider},
  {Huber}, {Pfeifer}, {Andreasen}, \& {Zechmeister}}]{Czesla2019}
{Czesla}, S., {Schr{\"o}ter}, S., {Schneider}, C.~P., {et~al.} 2019, {PyA:
  Python astronomy-related packages}.
\newblock \doeprint{1906.010}

\bibitem[{{Gaia Collaboration} {et~al.}(2018){Gaia Collaboration}, {Mignard},
  {Klioner}, {Lindegren}, {Hern{\'a}ndez}, {Bastian}, {Bombrun}, {Hobbs},
  {Lammers}, {Michalik}, {Ramos-Lerate}, {Biermann},
  {Fern{\'a}ndez-Hern{\'a}ndez}, {Geyer}, {Hilger}, {Siddiqui},
  {Steidelm{\"u}ller}, {Babusiaux}, {Barache}, {Lambert}, {Andrei}, {Bourda},
  {Charlot}, {Brown}, {Vallenari}, {Prusti}, {de Bruijne}, {Bailer-Jones},
  {Evans}, {Eyer}, {Jansen}, {Jordi}, {Luri}, {Panem}, {Pourbaix}, {Randich},
  {Sartoretti}, {Soubiran}, {van Leeuwen}, {Walton}, {Arenou}, {Cropper},
  {Drimmel}, {Katz}, {Lattanzi}, {Bakker}, {Cacciari}, {Casta{\~n}eda},
  {Chaoul}, {Cheek}, {De Angeli}, {Fabricius}, {Guerra}, {Holl}, {Masana},
  {Messineo}, {Mowlavi}, {Nienartowicz}, {Panuzzo}, {Portell}, {Riello},
  {Seabroke}, {Tanga}, {Th{\'e}venin}, {Gracia-Abril}, {Comoretto},
  {Garcia-Reinaldos}, {Teyssier}, {Altmann}, {Andrae}, {Audard},
  {Bellas-Velidis}, {Benson}, {Berthier}, {Blomme}, {Burgess}, {Busso},
  {Carry}, {Cellino}, {Clementini}, {Clotet}, {Creevey}, {Davidson}, {De
  Ridder}, {Delchambre}, {Dell'Oro}, {Ducourant}, {Fouesneau}, {Fr{\'e}mat},
  {Galluccio}, {Garc{\'\i}a-Torres}, {Gonz{\'a}lez-N{\'u}{\~n}ez},
  {Gonz{\'a}lez-Vidal}, {Gosset}, {Guy}, {Halbwachs}, {Hambly}, {Harrison},
  {Hestroffer}, {Hodgkin}, {Hutton}, {Jasniewicz}, {Jean-Antoine-Piccolo},
  {Jordan}, {Korn}, {Krone-Martins}, {Lanzafame}, {Lebzelter}, {L{\"o}ffler},
  {Manteiga}, {Marrese}, {Mart{\'\i}n-Fleitas}, {Moitinho}, {Mora}, {Muinonen},
  {Osinde}, {Pancino}, {Pauwels}, {Petit}, {Recio-Blanco}, {Richards},
  {Rimoldini}, {Robin}, {Sarro}, {Siopis}, {Smith}, {Sozzetti}, {S{\"u}veges},
  {Torra}, {van Reeven}, {Abbas}, {Abreu Aramburu}, {Accart}, {Aerts},
  {Altavilla}, {{\'A}lvarez}, {Alvarez}, {Alves}, {Anderson}, {Anglada Varela},
  {Antiche}, {Antoja}, {Arcay}, {Astraatmadja}, {Bach}, {Baker},
  {Balaguer-N{\'u}{\~n}ez}, {Balm}, {Barata}, {Barbato}, {Barblan}, {Barklem},
  {Barrado}, {Barros}, {Barstow}, {Bartholom{\'e} Mu{\~n}oz}, {Bassilana},
  {Becciani}, {Bellazzini}, {Berihuete}, {Bertone}, {Bianchi}, {Bienaym{\'e}},
  {Blanco-Cuaresma}, {Boch}, {Boeche}, {Borrachero}, {Bossini}, {Bouquillon},
  {Bragaglia}, {Bramante}, {Breddels}, {Bressan}, {Brouillet},
  {Br{\"u}semeister}, {Brugaletta}, {Bucciarelli}, {Burlacu}, {Busonero},
  {Butkevich}, {Buzzi}, {Caffau}, {Cancelliere}, {Cannizzaro}, {Cantat-Gaudin},
  {Carballo}, {Carlucci}, {Carrasco}, {Casamiquela}, {Castellani},
  {Castro-Ginard}, {Chemin}, {Chiavassa}, {Cocozza}, {Costigan}, {Cowell},
  {Crifo}, {Crosta}, {Crowley}, {Cuypers}, {Dafonte}, {Damerdji}, {Dapergolas},
  {David}, {David}, {de Laverny}, {De Luise}, {De March}, {de Souza}, {de
  Torres}, {Debosscher}, {del Pozo}, {Delbo}, {Delgado}, {Delgado}, {Diakite},
  {Diener}, {Distefano}, {Dolding}, {Drazinos}, {Dur{\'a}n}, {Edvardsson},
  {Enke}, {Eriksson}, {Esquej}, {Eynard Bontemps}, {Fabre}, {Fabrizio},
  {Faigler}, {Falc{\~a}o}, {Farr{\`a}s Casas}, {Federici}, {Fedorets},
  {Fernique}, {Figueras}, {Filippi}, {Findeisen}, {Fonti}, {Fraile}, {Fraser},
  {Fr{\'e}zouls}, {Gai}, {Galleti}, {Garabato}, {Garc{\'\i}a-Sedano},
  {Garofalo}, {Garralda}, {Gavel}, {Gavras}, {Gerssen}, {Giacobbe}, {Gilmore},
  {Girona}, {Giuffrida}, {Glass}, {Gomes}, {Granvik}, {Gueguen}, {Guerrier},
  {Guiraud}, {Guti{\'e}}, {Haigron}, {Hatzidimitriou}, {Hauser}, {Haywood},
  {Heiter}, {Helmi}, {Heu}, {Hofmann}, {Holland }, {Huckle}, {Hypki}, {Icardi},
  {Jan{\ss}en}, {Jevardat de Fombelle}, {Jonker}, {Juh{\'a}sz}, {Julbe},
  {Karampelas}, {Kewley}, {Klar}, {Kochoska}, {Kohley}, {Kolenberg},
  {Kontizas}, {Kontizas}, {Koposov}, {Kordopatis}, {Kostrzewa-Rutkowska},
  {Koubsky}, {Lanza}, {Lasne}, {Lavigne}, {Le Fustec}, {Le Poncin-Lafitte},
  {Lebreton}, {Leccia}, {Leclerc}, {Lecoeur-Taibi}, {Lenhardt}, {Leroux},
  {Liao}, {Licata}, {Lindstr{\o}m}, {Lister}, {Livanou}, {Lobel}, {L{\'o}pez},
  {Managau}, {Mann}, {Mantelet}, {Marchal}, {Marchant}, {Marconi}, {Marinoni},
  {Marschalk{\'o}}, {Marshall}, {Martino}, {Marton}, {Mary}, {Massari},
  {Matijevi{\v{c}}}, {Mazeh}, {McMillan}, {Messina}, {Millar}, {Molina},
  {Molinaro}, {Moln{\'a}r}, {Montegriffo}, {Mor}, {Morbidelli}, {Morel},
  {Morris}, {Mulone}, {Muraveva}, {Musella}, {Nelemans}, {Nicastro}, {Noval},
  {O'Mullane}, {Ord{\'e}novic}, {Ord{\'o}{\~n}ez-Blanco}, {Osborne}, {Pagani},
  {Pagano}, {Pailler}, {Palacin}, {Palaversa}, {Panahi}, {Pawlak},
  {Piersimoni}, {Pineau}, {Plachy}, {Plum}, {Poggio}, {Poujoulet},
  {Pr{\v{s}}a}, {Pulone}, {Racero}, {Ragaini}, {Rambaux}, {Regibo},
  {Reyl{\'e}}, {Riclet}, {Ripepi}, {Riva}, {Rivard}, {Rixon}, {Roegiers},
  {Roelens}, {Romero-G{\'o}mez}, {Rowell}, {Royer}, {Ruiz-Dern}, {Sadowski},
  {Sagrist{\`a} Sell{\'e}s}, {Sahlmann}, {Salgado}, {Salguero}, {Sanna},
  {Santana-Ros}, {Sarasso}, {Savietto}, {Schultheis}, {Sciacca}, {Segol},
  {Segovia}, {S{\'e}gransan}, {Shih}, {Siltala}, {Silva}, {Smart}, {Smith},
  {Solano}, {Solitro}, {Sordo}, {Soria Nieto}, {Souchay}, {Spagna}, {Spoto},
  {Stampa}, {Steele}, {Stephenson}, {Stoev}, {Suess}, {Surdej}, {Szabados},
  {Szegedi-Elek}, {Tapiador}, {Taris}, {Tauran}, {Taylor}, {Teixeira},
  {Terrett}, {Teyssand ier}, {Thuillot}, {Titarenko}, {Torra Clotet}, {Turon},
  {Ulla}, {Utrilla}, {Uzzi}, {Vaillant}, {Valentini}, {Valette}, {van Elteren},
  {Van Hemelryck}, {van Leeuwen}, {Vaschetto}, {Vecchiato}, {Veljanoski},
  {Viala}, {Vicente}, {Vogt}, {von Essen}, {Voss}, {Votruba}, {Voutsinas},
  {Walmsley}, {Weiler}, {Wertz}, {Wevers}, {Wyrzykowski}, {Yoldas},
  {{\v{Z}}erjal}, {Ziaeepour}, {Zorec}, {Zschocke}, {Zucker}, {Zurbach}, \&
  {Zwitter}}]{GAIA2018}
{Gaia Collaboration}, {Mignard}, F., {Klioner}, S.~A., {et~al.} 2018, \aap,
  616, A14, \dodoi{10.1051/0004-6361/201832916}

\bibitem[{{Gillon} {et~al.}(2013){Gillon}, {Triaud}, {Jehin}, {Delrez},
  {Opitom}, {Magain}, {Lendl}, \& {Queloz}}]{Gillon2013}
{Gillon}, M., {Triaud}, A.~H.~M.~J., {Jehin}, E., {et~al.} 2013, \aap, 555, L5,
  \dodoi{10.1051/0004-6361/201321620}

\bibitem[{{Hartman} \& {Bakos}(2016)}]{HartmanBakos2016}
{Hartman}, J.~D., \& {Bakos}, G.~{\'A}. 2016, Astronomy and Computing, 17, 1,
  \dodoi{10.1016/j.ascom.2016.05.006}

\bibitem[{{Howell} {et~al.}(2014){Howell}, {Sobeck}, {Haas}, {Still},
  {Barclay}, {Mullally}, {Troeltzsch}, {Aigrain}, {Bryson}, {Caldwell},
  {Chaplin}, {Cochran}, {Huber}, {Marcy}, {Miglio}, {Najita}, {Smith},
  {Twicken}, \& {Fortney}}]{Howell2014}
{Howell}, S.~B., {Sobeck}, C., {Haas}, M., {et~al.} 2014, \pasp, 126, 398,
  \dodoi{10.1086/676406}

\bibitem[{{Imamura} {et~al.}(2020){Imamura}, {Mitchell}, {Lebonnois}, {Kaspi},
  {Showman}, \& {Korablev}}]{Imamura2020}
{Imamura}, T., {Mitchell}, J., {Lebonnois}, S., {et~al.} 2020, \ssr, 216, 87,
  \dodoi{10.1007/s11214-020-00703-9}

\bibitem[{{Karalidi} {et~al.}(2016){Karalidi}, {Apai}, {Marley}, \&
  {Buenzli}}]{Karalidi2016}
{Karalidi}, T., {Apai}, D., {Marley}, M.~S., \& {Buenzli}, E. 2016, \apj, 825,
  90, \dodoi{10.3847/0004-637X/825/2/90}

\bibitem[{{Kellogg} {et~al.}(2017){Kellogg}, {Metchev}, {Heinze}, {Gagn{\'e}},
  \& {Kurtev}}]{Kellogg2017}
{Kellogg}, K., {Metchev}, S., {Heinze}, A., {Gagn{\'e}}, J., \& {Kurtev}, R.
  2017, \apj, 849, 72, \dodoi{10.3847/1538-4357/aa8e4f}

\bibitem[{{Kniazev} {et~al.}(2013){Kniazev}, {Vaisanen}, {Mu{\v{z}}i{\'c}},
  {Mehner}, {Boffin}, {Kurtev}, {Melo}, {Ivanov}, {Girard}, {Mawet},
  {Schmidtobreick}, {Huelamo}, {Borissova}, {Minniti}, {Ishibashi}, {Potter},
  {Beletsky}, {Buckley}, {Crawford}, {Gulbis}, {Kotze}, {Miszalski},
  {Pickering}, {Romero Colmenero}, \& {Williams}}]{Kniazev2013}
{Kniazev}, A.~Y., {Vaisanen}, P., {Mu{\v{z}}i{\'c}}, K., {et~al.} 2013, \apj,
  770, 124, \dodoi{10.1088/0004-637X/770/2/124}

\bibitem[{{Lang} {et~al.}(2010){Lang}, {Hogg}, {Mierle}, {Blanton}, \&
  {Roweis}}]{lang2010}
{Lang}, D., {Hogg}, D.~W., {Mierle}, K., {Blanton}, M., \& {Roweis}, S. 2010,
  \aj, 137, 1782

\bibitem[{{Libralato} {et~al.}(2016{\natexlab{a}}){Libralato}, {Bedin},
  {Nardiello}, \& {Piotto}}]{Libralato2016a}
{Libralato}, M., {Bedin}, L.~R., {Nardiello}, D., \& {Piotto}, G.
  2016{\natexlab{a}}, \mnras, 456, 1137, \dodoi{10.1093/mnras/stv2628}

\bibitem[{{Libralato} {et~al.}(2016{\natexlab{b}}){Libralato}, {Nardiello},
  {Bedin}, {Borsato}, {Granata}, {Malavolta}, {Piotto}, {Ochner}, {Cunial}, \&
  {Nascimbeni}}]{Libralato2016b}
{Libralato}, M., {Nardiello}, D., {Bedin}, L.~R., {et~al.} 2016{\natexlab{b}},
  \mnras, 463, 1780, \dodoi{10.1093/mnras/stw1932}

\bibitem[{{Limaye}(1986)}]{Limaye1986}
{Limaye}, S.~S. 1986, \icarus, 65, 335, \dodoi{10.1016/0019-1035(86)90142-9}

\bibitem[{{Lomb}(1976)}]{Scargle1976}
{Lomb}, N.~R. 1976, \apss, 39, 447, \dodoi{10.1007/BF00648343}

\bibitem[{{Luhman}(2013)}]{Luhman2013}
{Luhman}, K.~L. 2013, \apjl, 767, L1, \dodoi{10.1088/2041-8205/767/1/L1}

\bibitem[{{Mancini} {et~al.}(2015){Mancini}, {Giacobbe}, {Littlefair},
  {Southworth}, {Bozza}, {Damasso}, {Dominik}, {Hundertmark}, {J{\o}rgensen},
  {Juncher}, {Popovas}, {Rabus}, {Rahvar}, {Schmidt}, {Skottfelt}, {Snodgrass},
  {Sozzetti}, {Alsubai}, {Bramich}, {Calchi Novati}, {Ciceri}, {D'Ago},
  {Figuera Jaimes}, {Galianni}, {Gu}, {Harps{\o}e}, {Haugb{\o}lle}, {Henning},
  {Hinse}, {Kains}, {Korhonen}, {Scarpetta}, {Starkey}, {Surdej}, {Wang}, \&
  {Wertz}}]{Mancini2015}
{Mancini}, L., {Giacobbe}, P., {Littlefair}, S.~P., {et~al.} 2015, \aap, 584,
  A104, \dodoi{10.1051/0004-6361/201526899}

\bibitem[{{Marois} {et~al.}(2010){Marois}, {Zuckerman}, {Konopacky},
  {Macintosh}, \& {Barman}}]{Marois2010}
{Marois}, C., {Zuckerman}, B., {Konopacky}, Q.~M., {Macintosh}, B., \&
  {Barman}, T. 2010, \nat, 468, 1080, \dodoi{10.1038/nature09684}

\bibitem[{{Metchev} {et~al.}(2015){Metchev}, {Heinze}, {Apai}, {Flateau},
  {Radigan}, {Burgasser}, {Marley}, {Artigau}, {Plavchan}, \&
  {Goldman}}]{Metchev2015}
{Metchev}, S.~A., {Heinze}, A., {Apai}, D., {et~al.} 2015, \apj, 799, 154,
  \dodoi{10.1088/0004-637X/799/2/154}

\bibitem[{{Millar-Blanchaer} {et~al.}(2020){Millar-Blanchaer}, {Girard},
  {Karalidi}, {Marley}, {van Holstein}, {Sengupta}, {Mawet}, {Kataria}, {Snik},
  {Boer}, {Jensen-Clem}, {Vigan}, \& {Hinkley}}]{Millar-Blanchaer2020}
{Millar-Blanchaer}, M.~A., {Girard}, J.~H., {Karalidi}, T., {et~al.} 2020,
  \apj, 894, 42, \dodoi{10.3847/1538-4357/ab6ef2}

\bibitem[{{Nardiello} {et~al.}(2016{\natexlab{a}}){Nardiello}, {Libralato},
  {Bedin}, {Piotto}, {Borsato}, {Granata}, {Malavolta}, \&
  {Nascimbeni}}]{Nardiello2016b}
{Nardiello}, D., {Libralato}, M., {Bedin}, L.~R., {et~al.} 2016{\natexlab{a}},
  \mnras, 463, 1831, \dodoi{10.1093/mnras/stw2169}

\bibitem[{{Nardiello} {et~al.}(2016{\natexlab{b}}){Nardiello}, {Libralato},
  {Bedin}, {Piotto}, {Ochner}, {Cunial}, {Borsato}, \&
  {Granata}}]{Nardiello2016a}
---. 2016{\natexlab{b}}, \mnras, 455, 2337, \dodoi{10.1093/mnras/stv2439}

\bibitem[{{Nardiello} {et~al.}(2015){Nardiello}, {Bedin}, {Nascimbeni},
  {Libralato}, {Cunial}, {Piotto}, {Bellini}, {Borsato}, {Brogaard}, {Granata},
  {Malavolta}, {Marino}, {Milone}, {Ochner}, {Ortolani}, {Tomasella},
  {Clemens}, \& {Salaris}}]{Nardiello2015}
{Nardiello}, D., {Bedin}, L.~R., {Nascimbeni}, V., {et~al.} 2015, \mnras, 447,
  3536, \dodoi{10.1093/mnras/stu2697}

\bibitem[{{Nardiello} {et~al.}(2019){Nardiello}, {Borsato}, {Piotto},
  {Colombo}, {Manthopoulou}, {Bedin}, {Granata}, {Lacedelli}, {Libralato},
  {Malavolta}, {Montalto}, \& {Nascimbeni}}]{Nardiello2019}
{Nardiello}, D., {Borsato}, L., {Piotto}, G., {et~al.} 2019, \mnras, 490, 3806,
  \dodoi{10.1093/mnras/stz2878}

\bibitem[{{Porco} {et~al.}(2003){Porco}, {West}, {McEwen}, {Del Genio},
  {Ingersoll}, {Thomas}, {Squyres}, {Dones}, {Murray}, {Johnson}, {Burns},
  {Brahic}, {Neukum}, {Veverka}, {Barbara}, {Denk}, {Evans}, {Ferrier},
  {Geissler}, {Helfenstein}, {Roatsch}, {Throop}, {Tiscareno}, \&
  {Vasavada}}]{Porco2003}
{Porco}, C.~C., {West}, R.~A., {McEwen}, A., {et~al.} 2003, Science, 299, 1541,
  \dodoi{10.1126/science.1079462}

\bibitem[{{Radigan} {et~al.}(2012){Radigan}, {Jayawardhana}, {Lafreni{\`e}re},
  {Artigau}, {Marley}, \& {Saumon}}]{Radigan2012}
{Radigan}, J., {Jayawardhana}, R., {Lafreni{\`e}re}, D., {et~al.} 2012, \apj,
  750, 105, \dodoi{10.1088/0004-637X/750/2/105}

\bibitem[{{Ricker} {et~al.}(2014){Ricker}, {Winn}, {Vanderspek}, {Latham},
  {Bakos}, {Bean}, {Berta-Thompson}, {Brown}, {Buchhave}, {Butler}, {Butler},
  {Chaplin}, {Charbonneau}, {Christensen-Dalsgaard}, {Clampin}, {Deming},
  {Doty}, {De Lee}, {Dressing}, {Dunham}, {Endl}, {Fressin}, {Ge}, {Henning},
  {Holman}, {Howard}, {Ida}, {Jenkins}, {Jernigan}, {Johnson}, {Kaltenegger},
  {Kawai}, {Kjeldsen}, {Laughlin}, {Levine}, {Lin}, {Lissauer}, {MacQueen},
  {Marcy}, {McCullough}, {Morton}, {Narita}, {Paegert}, {Palle}, {Pepe},
  {Pepper}, {Quirrenbach}, {Rinehart}, {Sasselov}, {Sato}, {Seager},
  {Sozzetti}, {Stassun}, {Sullivan}, {Szentgyorgyi}, {Torres}, {Udry}, \&
  {Villasenor}}]{Ricker2014}
{Ricker}, G.~R., {Winn}, J.~N., {Vanderspek}, R., {et~al.} 2014, Society of
  Photo-Optical Instrumentation Engineers (SPIE) Conference Series, Vol. 9143,
  {Transiting Exoplanet Survey Satellite (TESS)}, 914320,
  \dodoi{10.1117/12.2063489}

\bibitem[{{Robinson} \& {Marley}(2014)}]{RobinsonMarley2014}
{Robinson}, T.~D., \& {Marley}, M.~S. 2014, \apj, 785, 158,
  \dodoi{10.1088/0004-637X/785/2/158}

\bibitem[{{Scargle}(1982)}]{Lomb1982}
{Scargle}, J.~D. 1982, \apj, 263, 835, \dodoi{10.1086/160554}

\bibitem[{{Showman} \& {Dowling}(2000)}]{Showman2000}
{Showman}, A.~P., \& {Dowling}, T.~E. 2000, Science, 289, 1737,
  \dodoi{10.1126/science.289.5485.1737}

\bibitem[{{Showman} \& {Kaspi}(2012{\natexlab{a}})}]{ShowmanKaspi2012}
{Showman}, A.~P., \& {Kaspi}, Y. 2012{\natexlab{a}}, ArXiv e-prints.
\newblock \doarXiv{1210.7573}

\bibitem[{{Showman} \& {Kaspi}(2012{\natexlab{b}})}]{Showman2013}
---. 2012{\natexlab{b}}, ArXiv e-prints.
\newblock \doarXiv{1210.7573}

\bibitem[{{Showman} \& {Kaspi}(2013)}]{ShowmanKaspi2013}
---. 2013, \apj, 776, 85, \dodoi{10.1088/0004-637X/776/2/85}

\bibitem[{{Showman} {et~al.}(2020){Showman}, {Tan}, \&
  {Parmentier}}]{Showman2020}
{Showman}, A.~P., {Tan}, X., \& {Parmentier}, V. 2020, arXiv e-prints,
  arXiv:2007.15363.
\newblock \doarXiv{2007.15363}

\bibitem[{{Showman} {et~al.}(2019){Showman}, {Tan}, \& {Zhang}}]{Showman2019}
{Showman}, A.~P., {Tan}, X., \& {Zhang}, X. 2019, \apj, 883, 4,
  \dodoi{10.3847/1538-4357/ab384a}

\bibitem[{{Simon} {et~al.}(2016){Simon}, {Rowe}, {Gaulme}, {Hammel},
  {Casewell}, {Fortney}, {Gizis}, {Lissauer}, {Morales-Juberias}, {Orton},
  {Wong}, \& {Marley}}]{Simon2016}
{Simon}, A.~A., {Rowe}, J.~F., {Gaulme}, P., {et~al.} 2016, \apj, 817, 162,
  \dodoi{10.3847/0004-637X/817/2/162}

\bibitem[{{Tan} \& {Showman}(2017)}]{Tan2017}
{Tan}, X., \& {Showman}, A.~P. 2017, \apj, 835, 186,
  \dodoi{10.3847/1538-4357/835/2/186}

\bibitem[{{Tan} \& {Showman}(2019)}]{Tan2019}
---. 2019, \apj, 874, 111, \dodoi{10.3847/1538-4357/ab0c07}

\bibitem[{{VanderPlas}(2018)}]{VanderPlas2018}
{VanderPlas}, J.~T. 2018, \apjs, 236, 16, \dodoi{10.3847/1538-4365/aab766}

\bibitem[{{Vos} {et~al.}(2017){Vos}, {Allers}, \& {Biller}}]{Vos2017}
{Vos}, J.~M., {Allers}, K.~N., \& {Biller}, B.~A. 2017, \apj, 842, 78,
  \dodoi{10.3847/1538-4357/aa73cf}

\bibitem[{{Yang} {et~al.}(2016){Yang}, {Apai}, {Marley}, {Karalidi}, {Flateau},
  {Showman}, {Metchev}, {Buenzli}, {Radigan}, {Artigau}, {Lowrance}, \&
  {Burgasser}}]{Yang2016}
{Yang}, H., {Apai}, D., {Marley}, M.~S., {et~al.} 2016, \apj, 826, 8,
  \dodoi{10.3847/0004-637X/826/1/8}

\bibitem[{{Zechmeister} \& {K{\"u}rster}(2009)}]{ZechmeisterKurster2009}
{Zechmeister}, M., \& {K{\"u}rster}, M. 2009, \aap, 496, 577,
  \dodoi{10.1051/0004-6361:200811296}

\bibitem[{{Zhang}(2020)}]{Zhang2020}
{Zhang}, X. 2020, Research in Astronomy and Astrophysics, 20, 099,
  \dodoi{10.1088/1674-4527/20/7/99}

\bibitem[{{Zhang} \& {Showman}(2014)}]{ZhangShowman2014}
{Zhang}, X., \& {Showman}, A.~P. 2014, \apjl, 788, L6,
  \dodoi{10.1088/2041-8205/788/1/L6}

\end{thebibliography}
\bibliographystyle{aasjournal}
\end{document}